\documentclass[a4paper,11pt]{article} 
\usepackage[latin1]{inputenc}
\usepackage[T1]{fontenc}
\usepackage{epsfig}
\usepackage{times}
\usepackage{amssymb}

\title{
Statistical Analysis for Long Term Correlations in the Stress Time Series
of Jerky Flow}
\author{D.~Kugiumtzis$^1$ and E.~C.~Aifantis$^2$ \\
$^1$ {\em Department of Mathematical, Physical and Computational Sciences,} \\
{\em Faculty of Engineering, Aristotle University of Thessaloniki,}\\
 {\em GR-54006, Thessaloniki, Greece} \\
$^2$ {\em Labaratory of Mechanics and Materials,} \\
 {\em Faculty of Engineering, Aristotle University of Thessaloniki,}\\
 {\em GR-54006, Thessaloniki, Greece}}
\date{}

\begin{document}
\maketitle

\newcommand{\ssub}[1]{\mbox{\tiny #1}}
\newcommand{\ssmall}[1]{\mbox{\scriptsize{#1}}}

\bibliographystyle{unsrt}

Key word index: {\em time series, nonlinearity, surrogate data test,
  PLC effect}

\begin{abstract}

Stress time series from the PLC effect typically exhibit stick-slips
of upload and download type. 
These data contain strong short-term correlations of a nonlinear type.  
We investigate whether there are also long term correlations, i.e.
the successive up-down patterns are generated by a deterministic 
mechanism.
A statistical test is conducted for the null 
hypothesis that the sequence of the up-down patterns is totally 
random.
The test is constructed by means of surrogate data, suitably 
generated to represent the null hypothesis. 
Linear and nonlinear estimates are used as test
statistics, namely autocorrelation, mutual information and Lyapunov 
exponents, which are found to have proper performance for the test.
The test is then applied to three stress time series under 
different experimental conditions.
Rejections are obtained for one of them and not with all 
statistics.
From the overall results we cannot conclude that the underlying mechanism 
to the PLC effect has long memory.

\end{abstract}

%**********************
\section{Introduction}
%**********************

The {\em Portevin-Le Ch\^{a}telier (PLC)\ effect} or jerky flow is one of 
the best studied forms of plastic instability in many metallic alloys when 
tensile specimens are deformed in a certain range of strain rates and 
temperatures. 
A distinct characteristic of PLC effect is the up-load 
and down-load behavior of the total stress vs time curves, due to the 
pinning / unpinning of lattice dislocations \cite{Bell73}.
As a result, the stress time series is comprised of successive {\em stick-slip}
patterns, i.e. slow rather linear up-trends followed by fast down-trends.
Simple physically-based mathematical models, suggested in the literature, 
reproduce partially this feature 
\cite{Aifantis87a,Aifantis88a,Bharathi02,Lebyodkin00,Hahn02}. 
Recently, data analysis using nonlinear methods give evidence for nonlinear
and chaotic behavior \cite{Ananthakrishna97,Ananthakrishna99,Noronha97}.

Nonlinear methods, mostly based on chaos theory, have been 
applied to real data from different fields with varying success 
\cite{Abarbanel96,Kantz97}.    
These methods provide estimates of dynamical characteristics of the
underlying system, such as topological or fractal dimenion, entropy
and Lyapunov exponent, as well as sophisticated data driven
models. 
However, the estimates are meaningful if there is evidence
that the underlying system is indeed nonlinear deterministic and 
eventually chaotic.  
Regarding the nonlinearity in the data, an indirect approach, 
namely testing the null hypothesis that the data are linear stochastic, 
has gained much interest in the last years.   
The test employs surrogate data to form the empirical distribution of
the test statistic under the null hypothesis
\cite{Theiler92,Schreiber99b,Kugiumtzis01a}. 

The methodology of nonlinear dynamics, including the surrogate data
test for nonlinearity has been applied recently to data from the PLC
effect. In a number of experiments of occurrence of the PLC effect,
nonlinear techniques were used aiming at characterising the
structure of the stress times series. Within certain experimental
range, jerky flow was reported to exhibit chaotic behaviour for
single crystals of Cu-Al alloys \cite{Ananthakrishna99,Lebyodkin00} as 
well as for Al-Mg polycrystals \cite{Noronha97,Ananthakrishna97}. 
Moreover, for both crystals, the surrogate data test gave evidence
for the existence of nonlinear dynamics. 
However, one could argue that this evidence is solely due to the presence 
of strong deterministic structure at small time scales within the upload 
or download phase, a quite obvious form of nonlinear dynamics.

In this work, we direct the statistical analysis to a different time scale.
We focus on correlations at a larger time scale that spans over the 
stick-slips, that is we investigate whether there is {\em long term} 
deterministic structure in addition to the {\em short term} nonlinear 
dynamics that forms the stick-slip patterns. 
The data analysis is done by means of hypothesis testing, where the 
null hypothesis is that the stick--slip sequence is random. 
For this we introduce an algorithm generating surrogate data with
the same stick-slip patterns at a random order and we apply several
linear and nonlinear statistics.
We use three stress time series from single crystals at different experimental 
conditions.

The outline of the paper is as follows. 
In Section\,\ref{Data}, the stress data are described and in 
Section\,\ref{Review}, the results of the standard nonlinear analysis 
on these are reviewed. 
In Section\,\ref{Test}, the surrogate test for the hypothesis
of random stick--slips is described, and in Section\,\ref{Results},
the results of the application of the test to the stress data are
presented. 
Finally, a discussion follows in Section\,\ref{Discussion}. 

%***************************************************
\section{The Stress Data}
\label{Data} 
%***************************************************

The stress time series are recordings of the total stress of single 
crystal Cu-10\% Al under compression at a constant strain rate
(these time series were used in \cite{Ananthakrishna95}).
The notation and some specifications for the data sets are given in 
Table\,\ref{datadetails}.
The selected records from the experiments regard plastic deformation
giving successive slow up-load and rapid down-load of stress.
% =========================================
Table 1 to be placed here
% =========================================

%***************************************************
\section{Review of the Nonlinear Analysis of the Stress Data}
\label{Review} 
%***************************************************

Recently, it has been shown that the stress time series at low and
medium strain rates are nonlinear and chaotic using standard nonlinear
methods based on chaos theory \cite{Ananthakrishna99,Lebyodkin00}. 
We start by reviewing these results on a particular time series, S1, 
described in Section\,\ref{Data}.

The stress time series and a segment of this is shown in
Fig.\,\ref{ser3} (top pannel). 
%====================================================
% Figures 1a and 1b to be placed here
%====================================================
The structure of successive stick--slip patterns for this stress range 
clearly indicates that the underlying system is deterministic at small 
time scales.
The other stress time series listed in Table\,\ref{datadetails} show 
the same feature of stick-slips.
There is no ambiguity at the level of sampling time as to the evolution of 
the up-load stress (stick phase); it is simply a linear upward trend. 
The same holds for the slip phase, which is much shorter and thus
the slope of the downward trend is very large.
This fine piecewise linear stress evolution cannot be generated by 
conventional stochastic Markov chain models, such as ARMA models,
neither by a linear deterministic system.  
It is therefore of no surprise that the estimates from 
nonlinear methods applied to this type of time series suggest nonlinear
deterministic structure (for a review on nonlinear methods refer to
\cite{Kugiumtzis94,Abarbanel96,Kantz97}).

The presence of nonlinear short-term dynamics can also be established 
statistically, testing the null hypothesis $\mbox{H}_0$ that the stress 
time series is generated by a linear (Gaussian) process, perturbed by 
a static, possibly nonlinear, transform 
\cite{Theiler92,Schreiber99b,Kugiumtzis01a}.
The transform is included in $\mbox{H}_0$ to explain deviations from 
Gaussian amplitude distribution of the data, which is often observed 
in real time series.
The test involves the generation of an ensemble of surrogate data,
i.e. time series that represent the null hypothesis, and the
computation of a test statistic $q$, here an estimate from a nonlinear
method, on the original and surrogate data. 
If the estimate $q_0$ on the original data does not lie within the 
empirical distribution of $q$ under $\mbox{H}_0$, formed by the 
estimates $q_1,q_2,\ldots,q_M$ on the $M$ surrogates, then $\mbox{H}_0$ 
is rejected and it is unlikely that the original time series is 
linear stochastic. 
The statistics $q_1,q_2,\ldots,q_M$ form typically a normal-like 
distribution.
Therefore the deviation of the statistic $q_0$ on the original data 
from the distribution of $q$ under $\mbox{H}_0$ is quantified by the
significance $S$ defined as
\begin{equation}
S = \frac{|q_0 - \bar{q}|}{s_q},
\label{eq:significance}
\end{equation}
where $\bar{q}$ and $s_q$ are the average and standard deviation of 
$q_1,q_2,\ldots,q_M$, respectively.
The rejection region for $\mbox{H}_0$ is formed by a lower limit for
$S$ given from the critical value of standard normal distribution at
a prespecified confidence level.
If $S>1.96$, $\mbox{H}_0$ is rejected at the $95\%$ confidence level.

One could easily discriminate the surrogate data (consistent to the 
abovementioned $\mbox{H}_0$) from the original stress data solely by 
eyeball judgement.
As shown in Fig.\,\ref{ser3} (middle panel), the surrogate time 
series fails to capture the special feature of the original data. 
The surrogate time series used in this Section are generated
using the STAP algorithm, recently presented in \cite{Kugiumtzis02}. 
The same results were obtained using the AAFT and IAAFT algorithms
(for a review on the algorithms and their performance see 
\cite{Kugiumtzis00}). 
The surrogate data are designed to mimic the original time series in
terms of amplitude distribution and autocorrelation and are otherwise
random.  
These two conditions are apparently not sufficient to preserve the 
stick--slip patterns of the stress data. 

In Fig.\,\ref{autmutllestap}, we show estimates of the
autocorrelation, mutual information and largest Lyapunov exponent on
the original stress data and $40$ STAP surrogate time series
(for review on these methods see 
\cite{Kugiumtzis94,Abarbanel96,Kantz97})\footnote{The algorithms in
the TISEAN software were used, see \cite{Hegger99b}.}.
%====================================================
% Figure 2a, 2b, and 2c to be placed here
%====================================================

The results on autocorrelation $r(\tau)$ for $\tau=1,\ldots,50\tau_s$, 
confirm that the STAP surrogate data have the same linear structure 
as the original time series 
The other two measures are both nonlinear. 
The mutual information $I(\tau)$ measures the general correlation,
linear and nonlinear.
The discrepancy in $I(\tau)$ for the original and the surrogate data,
shown in Fig.\,\ref{autmutllestap}b, suggests that the original data
contain nonlinear correlations and therefore give larger mutual
information for a long range of lags.  
The largest Lyapunov exponent $\lambda_1(m)$ measures the rate of 
divergence in the evolution of nearby trajectories in a reconstructed 
state space of dimension $m$.
Chaotic and stochastic systems have positive $\lambda_1$, and large
positive $\lambda_1$ indicate high complexity or stochasticity.  
As shown in Fig.\,\ref{autmutllestap}c, the original data obtain
significantly larger $\lambda_1(m)$ for the whole range of embedding
dimensions $m=1,\ldots,10$, which indicates that they exhibit more
complexity than the surrogate data.
For both nonlinear statistics the significance $S$ takes very high values
giving rejection of $\mbox{H}_0$ at essentially $100\%$ confidence level.
The same results were established with other generation algorithms for 
the surrogate data and other nonlinear estimates, i.e. correlation
dimension and fitting error of local averages. 

The nonlinear time series analysis we have done so far could reproduce 
a quite evident result, i.e. the stress time series contains nonlinear 
dynamics at small time scale. 
A more interesting question we investigate next is whether there is 
any evidence of determinism or correlation in the evolution of the
stick--slip patterns of the stress time series. 

%***************************************************
\section{Surrogate Data Test for Sequence of Patterns}
\label{Test} 
%***************************************************

We employ the statistical approach of surrogate data testing discussed 
in Section\,\ref{Review}, but the working $\mbox{H}_0$ now
is that the succession of the stick--slip patterns is random, i.e. the
stick--slip states are independent. 
The surrogate data for this $\mbox{H}_0$ should have the same
stick--slip structure as the stress data, but at a random order. 

% -------------------------------
\subsection{The SUDT algorithm}
% -------------------------------

We have built an algorithm, called {\em Stochastic Up-Down Trends}
(SUDT), to generate the surrogate data for this $\mbox{H}_0$.
The algorithm permutates randomly the stick--slips of the original 
time series taking care that the range of the original data is
retained. Specifically, the steps of the algorithm are as follows:
\begin{enumerate}
  \item Scan the original time series $\mathbf{x}=[x_1,x_2,\ldots,x_n]'$, 
    identify and store the up-down
    trend patterns, as well as the global minimum $x_{\ssmall{min}}$, 
    the global maximum $x_{\ssmall{max}}$, the smallest end-point of the 
    up-trend $x_u$ and the largest end point of the down-trend $x_d$.
  \item The surrogate time series $\mathbf{z}$ to be constructed starts 
    at the same data point as $\mathbf{x}$, i.e. $z_1 = x_1$ (a random 
    point could be chosen as well). 
  \item Using discrete uniform distribution, draw randomly an up-down
    trend segment from those stored in step 1, and displace it so that its
    starting point coincides with the currently last point of the time
    series $\mathbf{z}$ (for the first iteration this is $z_1$).
    \label{step3}
  \item Check whether the ``up'' end-point of the chosen up-down trend
    segment is between $x_u$ and $x_{\ssmall{max}}$, and the ``down'' 
    end-point is between $x_{\ssmall{min}}$ and $x_d$.
    If the two end-points are within the given limits, then accept the
    up-down trend (the ``down'' end-point of the accepted trend is now
    the last data point of the time series).
    If not, then discard the up-down trend and repeat step~\ref{step3}. 
  \item Repeat the last two steps until the time series $\mathbf{z}$ is 
    as long as the original time series (eventually truncating the last
    eligible trend). 
\end{enumerate}

Note that the algorithm assumes that the original time series starts
with an upward trend. 
We adjust accordingly the data sets by dropping a few samples from the 
beginning if necessary.
Alternatively, one can simply reverse the magnitudes of the
original data (e.g. multiply by -1) before applying the algorithm.

This algorithm implements bootstrapping on blocks of data, i.e. the
up-down trends from the original time series, allowing repetitions of 
the same block of data in the surrogate time series. 
Simple shuffling of the original trends cannot be done because the
end-points have to match. 

It was found necessary to constrain the random selection of the
up-down trends using lower and upper limits for both the ``up'' and
``down'' end-points of each trend in order to keep the generated 
surrogate time series $\mathbf{z}$ within the bounds of the original
data. 
The use of the global minimum and maximum ($x_{\ssmall{min}}$ and 
$x_{\ssmall{max}}$) alone led to edge effect problems, so the additional 
limits of $x_u$ and $x_d$ had to be introduced to assure robust 
execution of the algorithm.
This increases the frequency of discarding randomly selected up-down
trends and mars the random order of the up-down trends.

% ---------------------------------------
\subsection{Examples with simulated data}
% ---------------------------------------

The surrogate data generated by the SUDT algorithm represent the 
$\mbox{H}_0$ of independent stick--slip states in the time series. 
Certainly, the stick--slip states of the original data may be
correlated implying that the underlying mechanism exhibits a
deterministic structure on longer time scales, which is an interesting
possible aspect for the PLC effect. 

Using simulated data, we show that the standard methods of time series
analysis have actually discriminative power and can distinguish the 
original time series from its SUDT surrogates only when this is the case. 
For this, we use deterministic and stochastic time series of the 
stick-slip type.
For the deterministic case, we use $2000$ data of the
log--transformed $w$ variable of the R\"{o}ssler hyperchaos system
\cite{Roessler79}, sampled at time $\tau_s=0.1\mbox{sec}$, call it
$\mathbf{x}^d$. 
This time series exhibits stick--slip structure.
For the stochastic case (time series with random stick--slip states),
we simply use a time series derived by the SUDT algorithm on
$\mathbf{x}^d$, call it $\mathbf{x}^s$.   
Two segments of the two time series are shown in Fig.\,\ref{lhy1}a.
%====================================================
% Figures 3a and 3b to be placed here
%====================================================
Note that the time series $\mathbf{x}^d$ and $\mathbf{x}^s$ have
the same structure and cannot be distinguished by eyeball judgement.
In the generation of 40 SUDT surrogates, there were on average 
48 rejections of candidate stick-slips for $\mathbf{x}^d$ (which
is comprised of 193 stick--slip patterns) and about the same for
$\mathbf{x}^s$, so that the shuffling can indeed be considered random.
The SUDT surrogate data for each of the two time series possess
similar amplitude distribution to the original ones, as shown in 
Fig.\,\ref{lhy1}b.
The same holds for the distribution of the up and down velocities and
for the distribution of the up and down times.
The preservation of all these distributions signifies the successful
performance of the SUDT algorithm.

We apply the linear and nonlinear test statistics on $\mathbf{x}^d$, 
$\mathbf{x}^s$ and their respective SUDT surrogates.
The results are shown in Fig.\,\ref{lhy2}.
%====================================================
% Figures 4a, 4b, 4c, 4d, 4e and 4f to be placed here
%====================================================
The autocorrelation $r(\tau)$ does not discriminate $\mathbf{x}^s$ 
from its SUDT surrogates, as expected, but the same holds for 
$\mathbf{x}^d$ (see Fig.\,\ref{lhy2}a).
Specifically for $\mathbf{x}^d$, rejection of $\mbox{H}_0$ at the $95\%$ 
confidence level could only be established for a small range of delays 
around $\tau=5$, as shown in Fig.\,\ref{lhy2}d.
So, for the chaotic time series from R\"{o}ssler hyperchaos the linear
test statistic has essentially no discriminative power.
Note that there is no reason to believe that this is always the case 
with deterministic systems.
However, the same results were obtained also on a quasi-periodic system 
with stick--slip structure (a two torus, for description of the system
see \cite{Fraser89}).    

The chaotic deterministic data $\mathbf{x}^d$ are correctly
distinguished from their respective SUDT surrogates with both nonlinear 
statistics, i.e. the mutual information $I(\tau)$ and the
largest Lyapunov exponent $\lambda_1(m)$, as shown in Fig.~\ref{lhy2}b 
and c.
Subsequently, the $\mbox{H}_0$ of independent stick--slip states is
rejected at the $95\%$ confidence levels for a long range of the free
parameter of its statistic (see Fig.~\ref{lhy2}e and f). 
On the other hand, the $\mathbf{x}^s$ data are correctly
not distinguished from their respective SUDT surrogates by either
$\lambda_1(m)$ or $I(\tau)$, and $\mbox{H}_0$ is not rejected for any
value of the free parameter of the two statistics.
Very similar results were obtained on the quasi-periodic system (where
$S$ obtained larger values for the deterministic system) as well as 
when using other nonlinear statistics, e.g. local average maps and 
entropy.

These findings show that even standard nonlinear statistics that are
not tailored for this particular test can distinguish correctly 
correlated stick--slip states from non-correlated stick--slip
states of similar shape. 

%***************************************************
\section{Application of the Test to the Stress Data}
\label{Results} 
%***************************************************

In order to avoid false local minima and maxima the stress time series 
were smoothed using a finite impulse response filter prior to the 
identification of the stick--slips.
Note that there was no further use of the smoothed time series and the 
shuffling in the SUDT algorithm was done on the original stick--slips.
The SUDT surrogate time series preserve well the stick--slip patterns
of all the PLC stress time series.
In Fig.~\ref{ser3}, this is shown for the stress time series S1 
and one SUDT surrogate of this (upper and lower pannel, respectively).

For the surrogate data to be proper for the test, we require good match 
of the distribution of the data, the distribution of the velocity of 
the up and down trends, as well as the distribution of the time of the 
up and down trends.
The velocity and time distributions were preserved in the SUDT surrogates
for all three stress time series.
The amplitude distribution was well preserved for S2, sufficiently preserved
for S1 and not preserved for S3, as shown in Fig.~\ref{S1S2S3amplitude}.
%====================================================
% Figure 5a, 5b and 5c to be placed here
%====================================================
It turns out that the stick--slip time series generated by the SUDT 
algorithm tend to possess symmetric amplitude distribution, so that 
when the original data have skewed distribution (as is the case with S3)
deviations in amplitude distribution do occur (see 
Fig.~\ref{S1S2S3amplitude}c).
This constitutes a shortcoming of the SUDT algorithm to provide proper 
stick--slip surrogates.
So, whenever the amplitude distribution is not preserved one may question 
the outcome of the test as deviations in the test statistics that may
lead to rejection of $\mbox{H}_0$ may be assigned to the mismatch
of amplitude distribution.

In Fig.~\ref{S1S2S3autmutlya}, the outcome of the test using the three
test statistics is shown.
The linear statistic $r(\tau)$ distinguishes the time series S1 from
its 40 surrogates (for a long range of $\tau$), but not S2.
For S3, the $r(\tau)$ for the SUDT surrogates is much higher (accordingly, 
$S$ takes very high values not shown in Fig.~\ref{S1S2S3autmutlya}d for
$\tau \leq 13$), but this mismatch may be due to the lack of match in 
amplitude distribution, so it cannot be regarded as genuine discrimination 
that would correspond to rejection of $\mbox{H}_0$.
Therefore, the clear rejection for S3 also with the nonlinear statistics 
are not reliable. 
%====================================================
% Figures 6a, 6b, 6c, 6d, 6e and 6f to be placed here
%====================================================

It should be noted that for S1, the original $r(\tau)$ for $\tau<40$ is 
actually smaller in amplitude than for the surrogates suggesting the 
opposite of the alternative hypothesis we attempt to establish, i.e. the
surrogate data involve more correlations than the original data.
A possible explanation for this would be the discrepancy at the bulk
of the amplitude distribution between S1 and its surrogates (see 
Fig.~\ref{S1S2S3amplitude}a).
We note also that sucl long range correlations are often due to drift in 
the data. 

The $I(\tau)$ statistic that measures both linear and nonlinear 
correlations is at the same level for S1 and its SUDT surrogates,
as shown in Fig.~\ref{S1S2S3autmutlya}b.
Compared to the results with $r(\tau)$, it seems that S1 contains 
nonlinear correlations not present in the SUDT surrogates.
However, $S$ is below the threshold for rejection of $\mbox{H}_0$ at
the $95\%$ confidence level for all but very small $\tau$
(see Fig.~\ref{S1S2S3autmutlya}e).
The $\lambda_1(m)$ statistic shows also a difference in nonlinear 
correlations between S1 and its surrogates giving confident rejections
at the $95\%$ level for $m>2$, as shown in Fig.~\ref{S1S2S3autmutlya}c
and f.

The stress time series S2 could not be discriminated from its
SUDT surrogates with any of the three statistics, indicating that 
it has no correlations between stick--slips.

The results on the three stress time series suggest that there is not 
enough statistical evidence to establish that the stick--slip states of 
the stress time series from plastic deformation of single crystals are 
correlated and thus that there is a deterministic system at large time 
scales that controls the evolution of the stick--slip states.

%***************************************************
\section{Discussion}
\label{Discussion} 
%***************************************************

It is plausible that the evolution of the stress in the PLC effect is 
rather deterministic at short time scales as the stress time series 
has a characteristic stick-slip structure. 
We employed statistical testing to investigate whether there is a 
deterministic mechanism that controls the stress at larger time scales 
that span over the duration of the stick-slip states.
The standard surrogate data test for nonlinearity, used recently to 
establish determinism and nonlinearity for the PLC effect, is not 
suitable for the question of interest as the surrogate data do not 
preserve the stick-slip structure.
Subsequently, the disrcimination between original stress time series 
and surrogate data (generated under the null hypothesis of linear 
stochastic system) is guaranteed. 
However, this result does not establish the presence of correlations 
between the stick--slips that should be present if the underlying
mechanism is nonlinear deterministic.
Indeed, our simulations showed that time series comprised of 
uncorrelated stick--slips are also discriminated from this type of 
surrogates, questioning the appropriateness of the surrogate data test
for nonlinearity for this type of time series.  

We designed the SUDT algorithm to generate surrogate data of stick--slip 
structure and performed the surrogate data test for the null hypothesis 
of independent stick-slip states.
Nonlinear statistics as the ones used for the test for nonlinearity 
turned out to perform appropriately when applied to simulated data.
However, the power of the statistics on chaotic time series with stick-slip
structure was not as high as for the quasi-periodic systems.

We applied this test on three stress time series from plastic deformation
of single crystal Cu-10\% Al under compression at different strain rates.
We used one linear test statistic, i.e. the autocorrelation, and two 
nonlinear statistics: the mutual information and the largest Lyapunov 
exponent.
The null hypothesis could be rejected, but not clearly, for the stress time 
series obtained at low constant strain rate 
($\dot{\epsilon}=3.3\cdot10^{-6}\mbox{s}^{-1}$) and could not be rejected
with any test statistic for the stress time series obtained at medium 
constant strain rate ($\dot{\epsilon}=37\cdot10^{-6}\mbox{s}^{-1}$).
For the third time series at larger constant strain rate 
($\dot{\epsilon}=107\cdot10^{-6}\mbox{s}^{-1}$), the SUDT algorithm failed
to match the amplitude distribution and thus the rejection obtained 
with the test statistics is questioned.
Overall, the statistical testing could not establish that the stress 
time series contain significant correlations. 
However, this is a pilot study and a systematic application of the
test to stress time series under varying experimental conditions is
planned.

An improvement of the SUDT algorithm would be to constraint the surrogates 
to match the amplitude distribution of the original time series, but there 
does not seem to be an obvious way to do this. 
Our simulations showed that the match is maintained through the suffling
of the stick--slips, but for one stress time series this failed. 
Also, the test may be improved by employing other test statistics that
are tailored to capture the information relating the stick-slip states,
such as correlation between the lengths of successive up-down trends or  
between the magnitudes of successive turning points.

%=========================
\section*{Acknowledgements}
%=========================

Partial support of the European Community's Human Potential
Programme under contract HPRN-CT-2002-00198 (DEFINO) and research
interaction between AUTh/Greece and TUB/Germany are 
acknowledged. 

%====================================================
%       And the bibliography, Using BibTeX
% \clearpage
% \bibliography{$HOME/LaTeX/ChaosReferences} 
%====================================================

\clearpage

% ====== Table 1 =========================================
\begin{table}[htb]
\centerline{\begin{tabular}{||c|c|c|c|c||} \hline\hline
notation & T$[^o\mbox{C}]$ & $\dot{\epsilon}[10^{-6}\mbox{s}^{-1}]$ & 
$\tau_s[\mbox{s}]$ & $T[\mbox{s}]$ \\ \hline
S1 &   & $3.3$ & 0.15 & 1000 \\
S2 & 300 & $37$ & 0.006 & 30.28 \\
S3 & 300 & $107$ & 0.06 & 400.02 \\ \hline\hline
\end{tabular}}
\caption{Notation and specification of the stress time series. 
In the second column is the temperature (T), in the third column the 
strain rate ($\dot{\epsilon}$), in the fourth column the sampling time 
($\tau_s$) and in the last column is the recording time ($T$).}
\label{datadetails}
\end{table}
% ========================================================

\clearpage
% ======== Figure 1 ======================================
\begin{figure}[htb] % 1
\caption{(a) The stress time series, a STAP surrogate and a SUDT
  surrogate, from top to bottom. 
The two gray vertical lines denote the segments of the data, which are
enlarged in (b).}
\label{ser3}
\end{figure}

%========= Figure 2 =======================================
\begin{figure}[htb] % 2
\caption{Estimates on the stress time series and $40$ STAP surrogates:
in (a) autocorrelation $r(\tau)$ vs lag $\tau$, in (b) mutual 
information $I(\tau)$ vs $\tau$ and in (c) largest Lyapunov exponent 
$\lambda_1$ vs embedding dimension $m$.} 
\label{autmutllestap}
\end{figure}
%====================================================

%========= Figure 3========================================
\begin{figure}[htb] % 1
\caption{(a) A segment of the time series of the R\"{o}ssler hyperchaos
  system $\mathbf{x}^d$ (upper panel) and a segment of a SUDT
  surrogate of it $\mathbf{x}^s$ (bottom panel).  
(b) The amplitude distribution (histogram) of $\mathbf{x}^d$ and $40$
SUDT surrogates (upper panel), and of $\mathbf{x}^s$ and $40$ SUDT
surrogates (lower panel). 
Black thick lines denote the original data and gray lines denote the
surrogates.} 
\label{lhy1}
\end{figure}
%====================================================

%========== Figure 4 =========================================
\begin{figure}[htb] 
\caption{(a) The autocorrelation $r(\tau)$ of $\mathbf{x}^d$ and $40$
SUDT surrogates (upper panel), and of $\mathbf{x}^s$ and $40$ SUDT
surrogates (lower panel). 
(b) The mutual information $I(\tau)$ for the same sets of data as for (a).
(c) The largest Lyapunov exponent $\lambda_1(m)$ for the same sets of
data as for (a).
(d) The significance $S$ for $r(\tau)$ in (a).
(e) $S$ for $I(\tau)$ in (b).
(f) $S$ for $\lambda_1(m)$ in (c).
For (a), (b) and (c), the black thick lines denote the original data and 
the gray lines denote the surrogates.
For (d), (e) and (f), the level of $S=1.96$ is denoted by a horizontal gray
line.} 
\label{lhy2}
\end{figure}
%====================================================

%========== Figure 5 ==========================================
\begin{figure}[htb] 
\caption{Amplitude distribution of the three stress time series and 
their respective $40$ SUDT surrogates: (a) S1, (b) S2, (c) S3.} 
\label{S1S2S3amplitude}
\end{figure}
%====================================================

%========== Figure 6 =========================================
\begin{figure}[htb]
\caption{(a) The autocorrelation $r(\tau)$ of the stress time series S1,
S2, S3, and their $40$ SUDT surrogates at the upper, middle and lower panel,
respectively.
(b) The mutual information $I(\tau)$ for the same sets of data as for (a).
(c) The largest Lyapunov exponent $\lambda_1(m)$ for the same sets of
data as for (a).
(d) The significance $S$ for $r(\tau)$ in (a).
(e) $S$ for $I(\tau)$ in (b).
(f) $S$ for $\lambda_1(m)$ in (c).
For (a), (b) and (c), the black thick lines denote the original data and 
the gray lines denote the surrogates.
For (d), (e) and (f), the level of $S=1.96$ is denoted by a horizontal gray
line.} 
\label{S1S2S3autmutlya}
\end{figure}
%====================================================

\clearpage
\centerline{Figure 1a}
\psfig{file=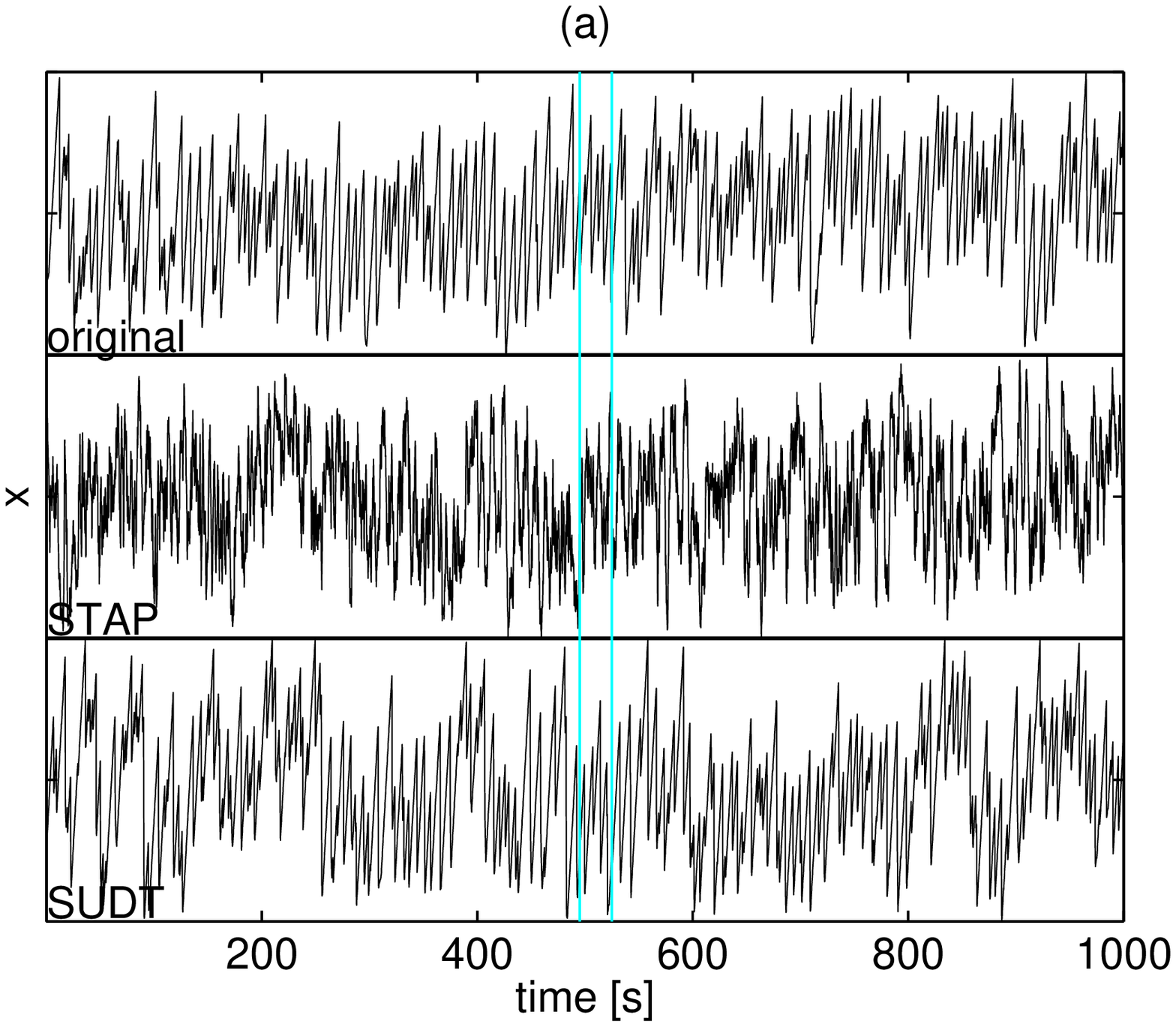,height=10cm,width=10cm}

\clearpage
\centerline{Figure 1b}
\psfig{file=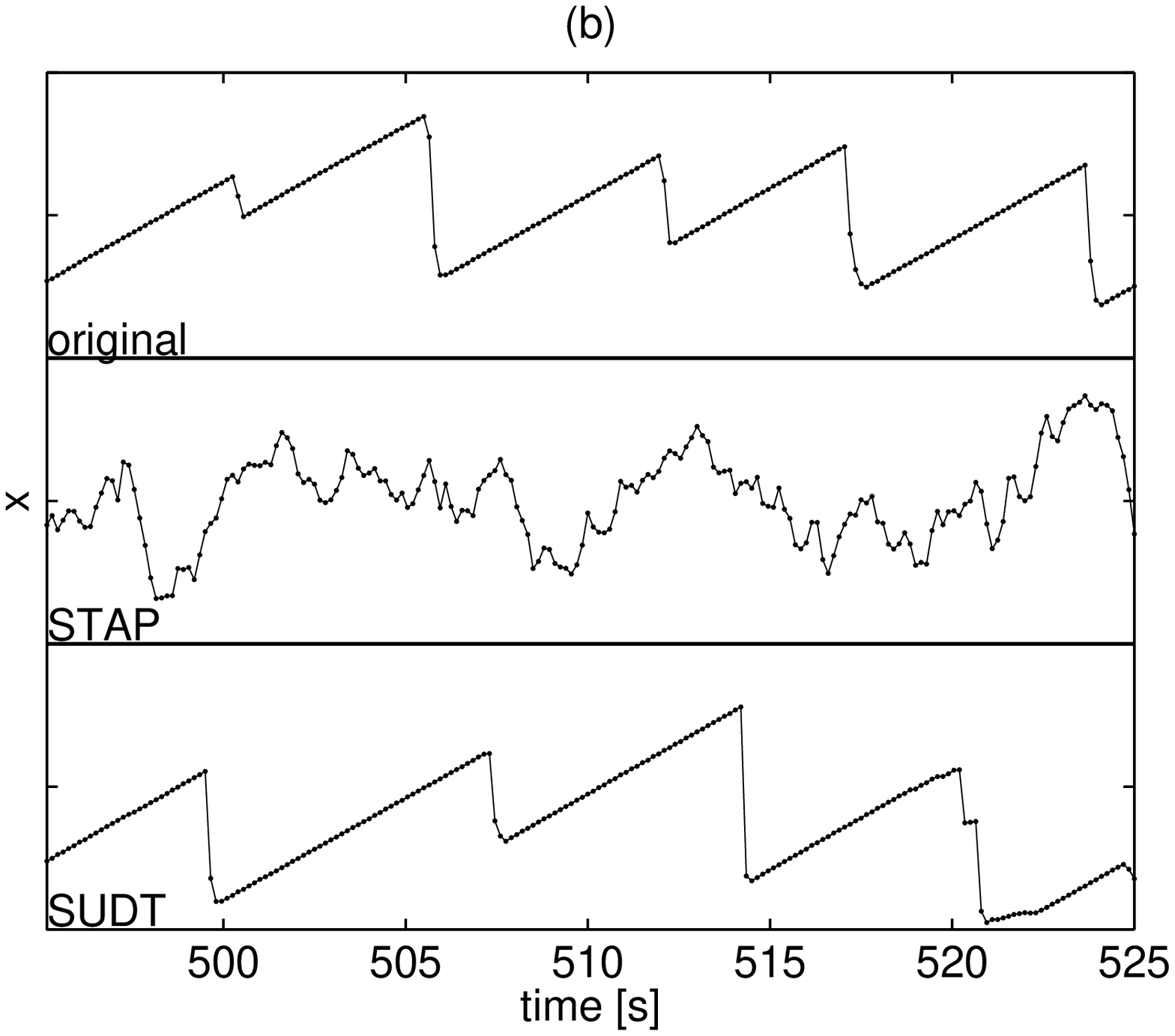,height=10cm,width=10cm}

\clearpage
\centerline{Figure 2a}
\psfig{file=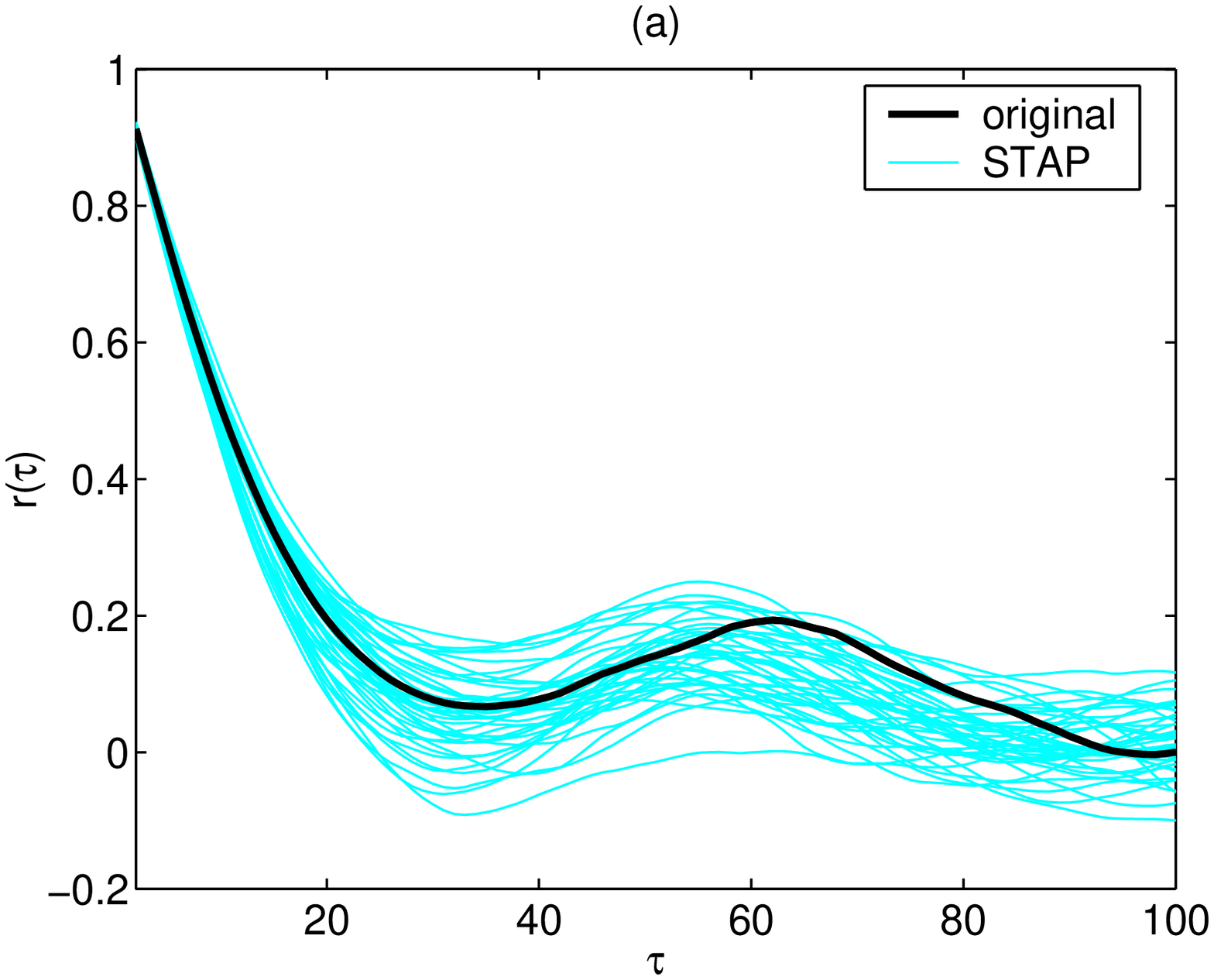,height=10cm,width=10cm}

\clearpage
\centerline{Figure 2b}
\psfig{file=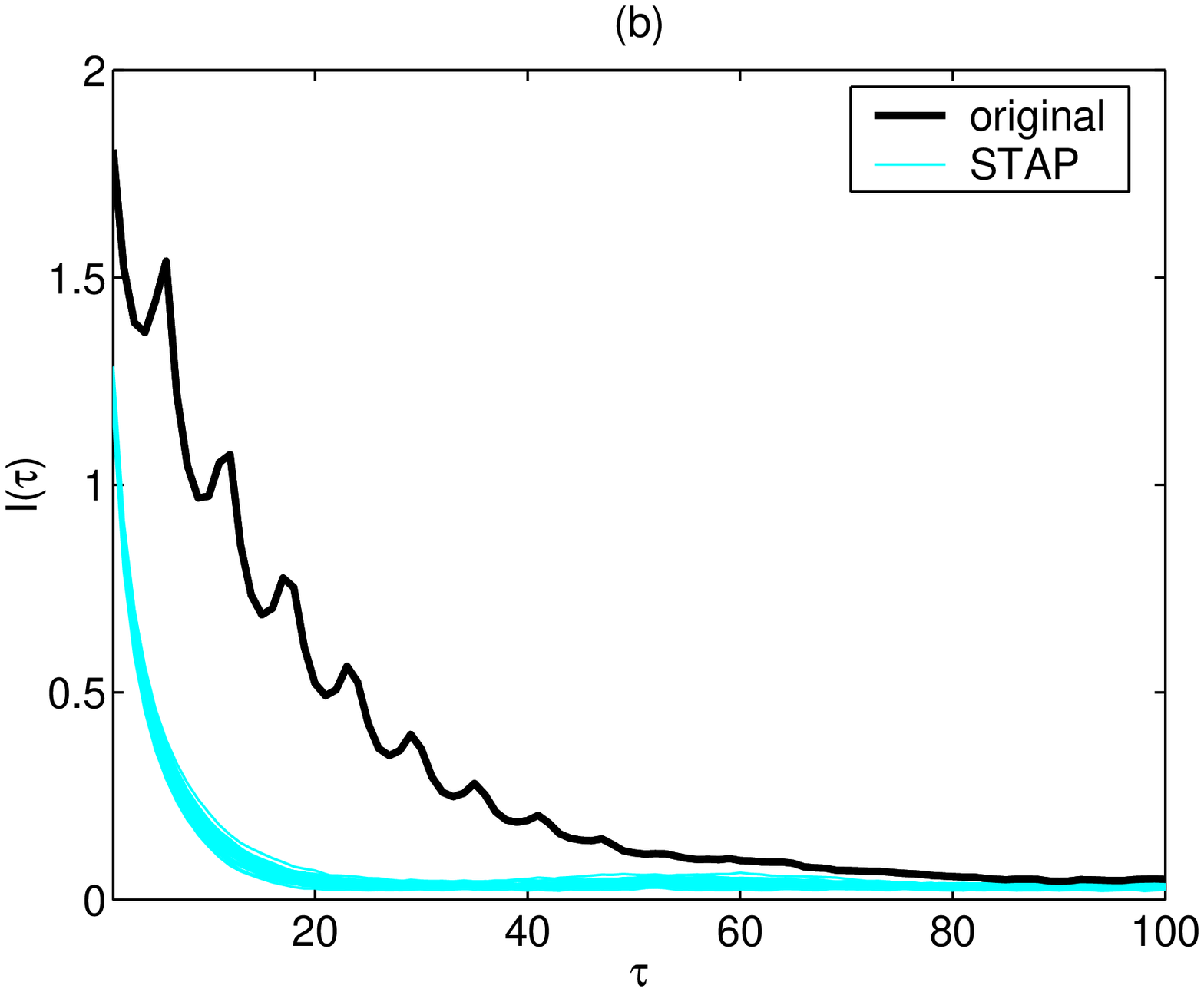,height=10cm,width=10cm}

\clearpage
\centerline{Figure 2c}
\psfig{file=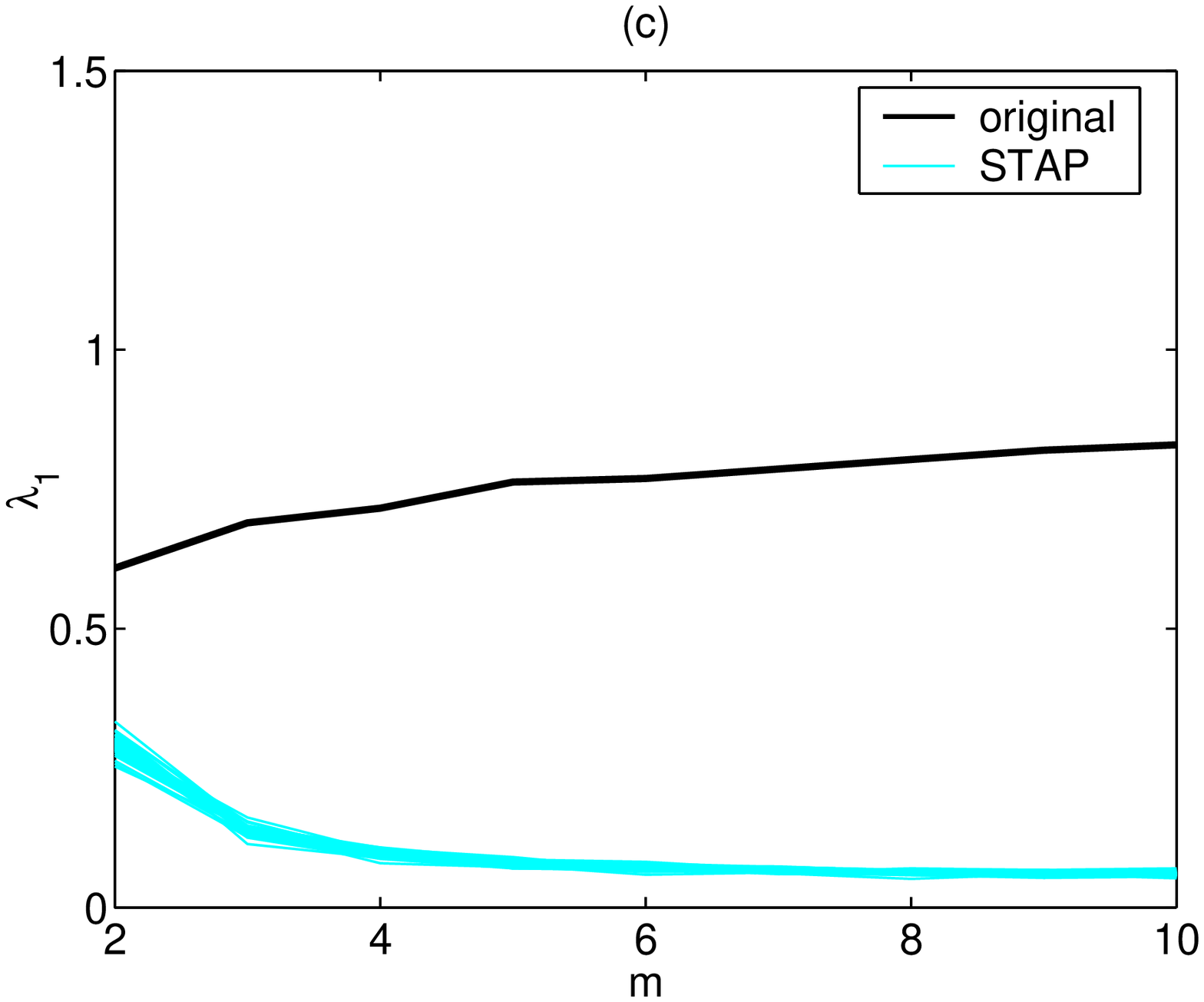,height=10cm,width=10cm}

\clearpage
\centerline{Figure 3a}
\psfig{file=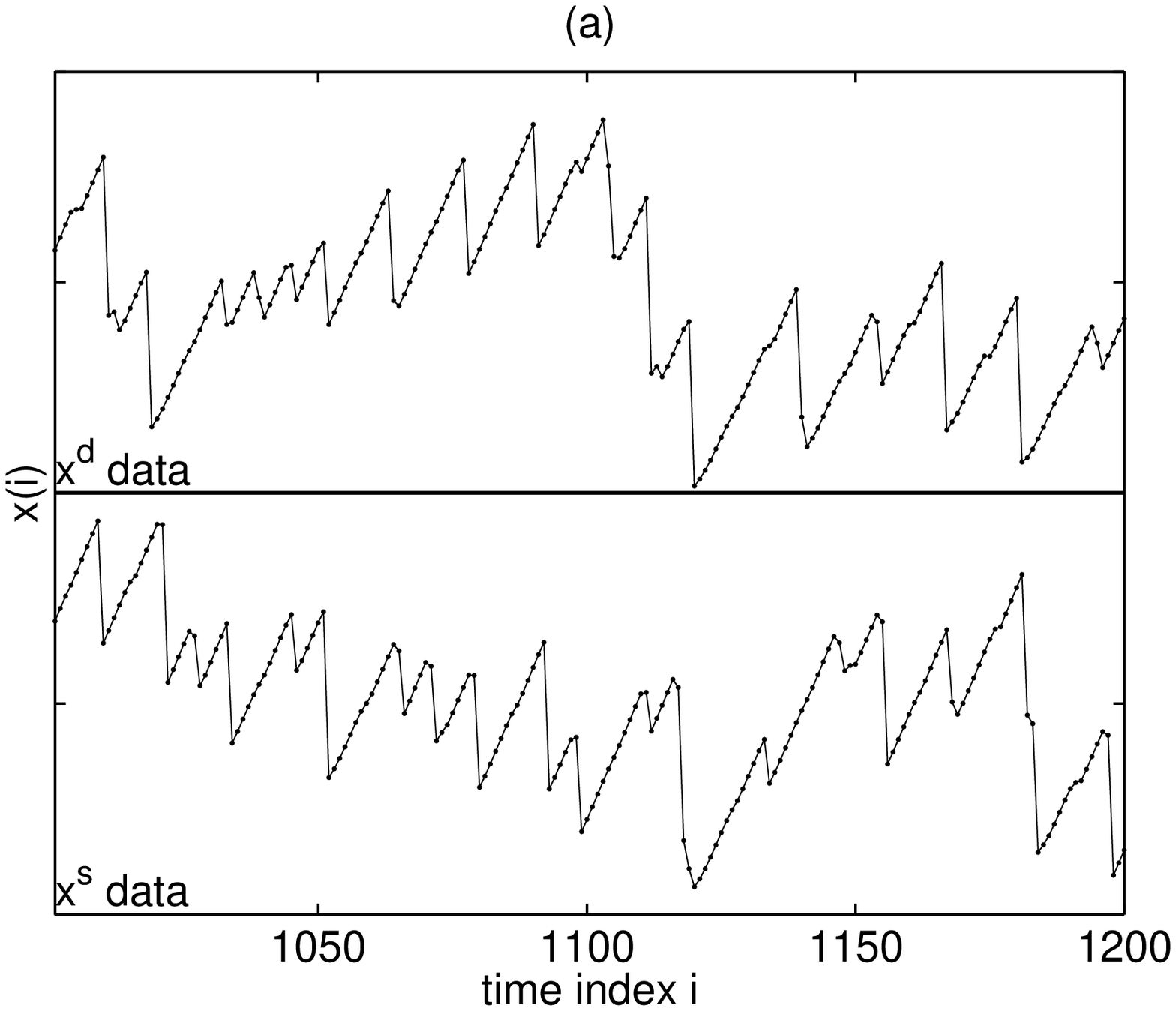,height=10cm,width=10cm}

\clearpage
\centerline{Figure 3b}
\psfig{file=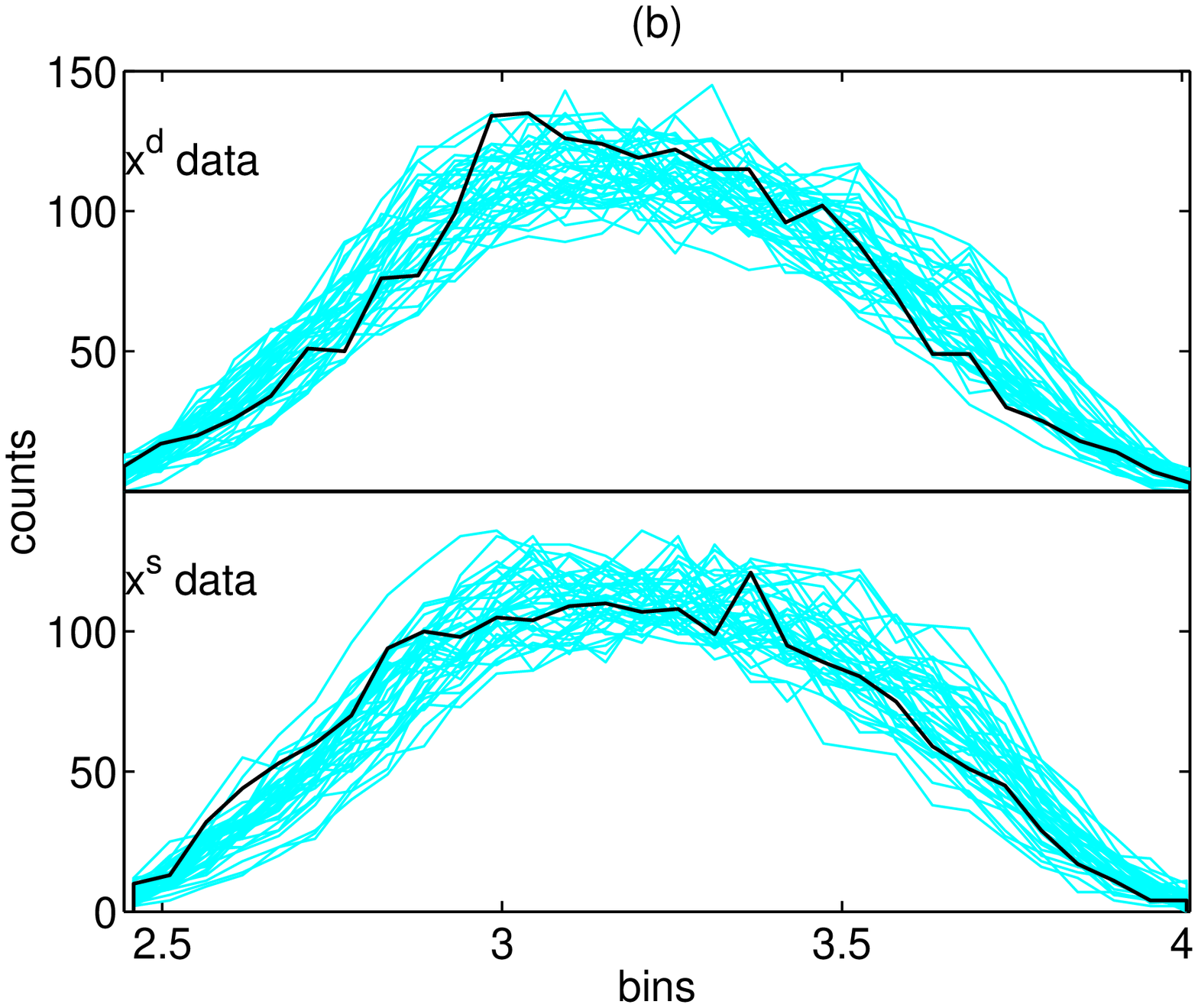,height=10cm,width=10cm}

\clearpage
\centerline{Figure 4a}
\psfig{file=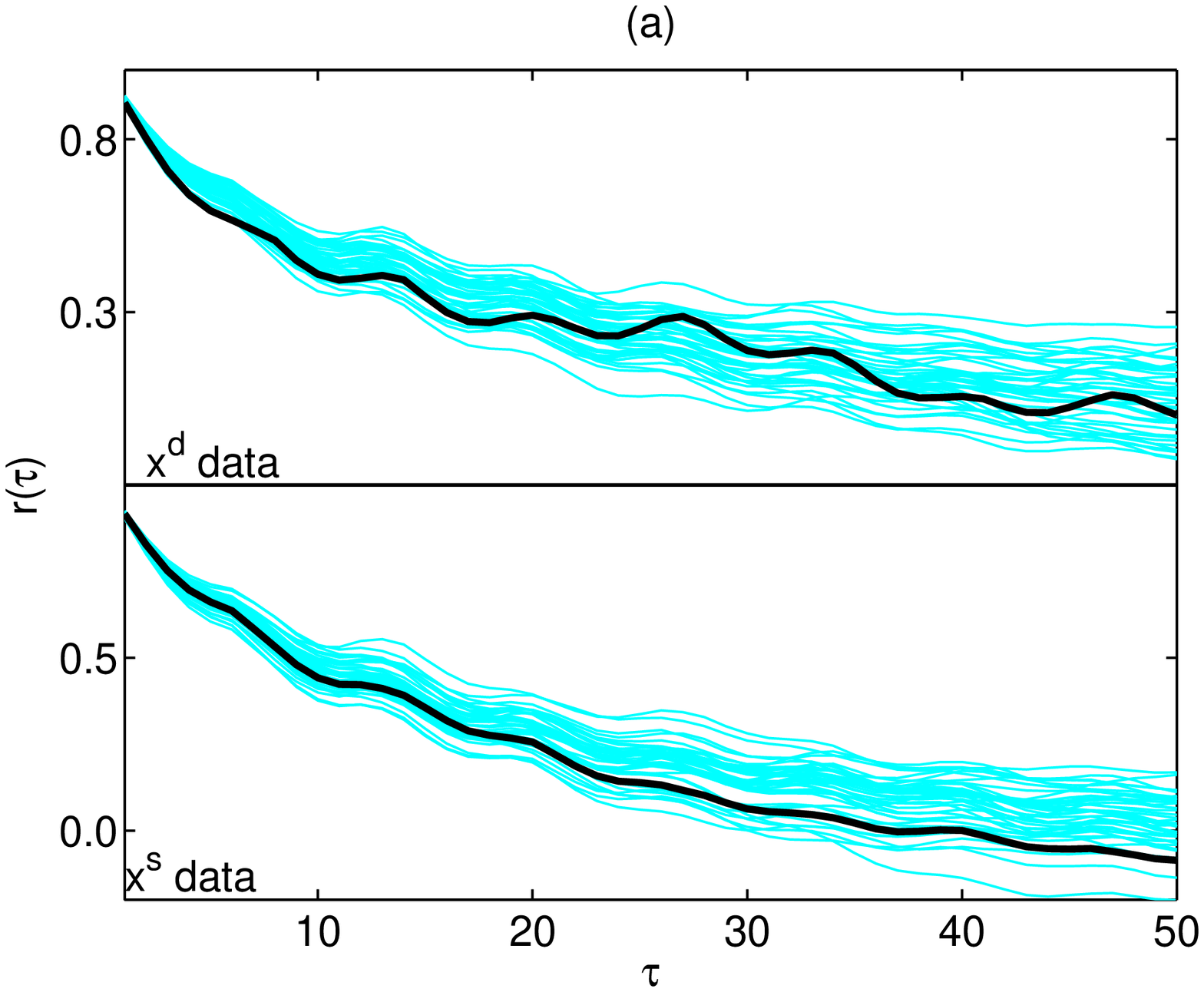,height=10cm,width=10cm}

\clearpage
\centerline{Figure 4b}
\psfig{file=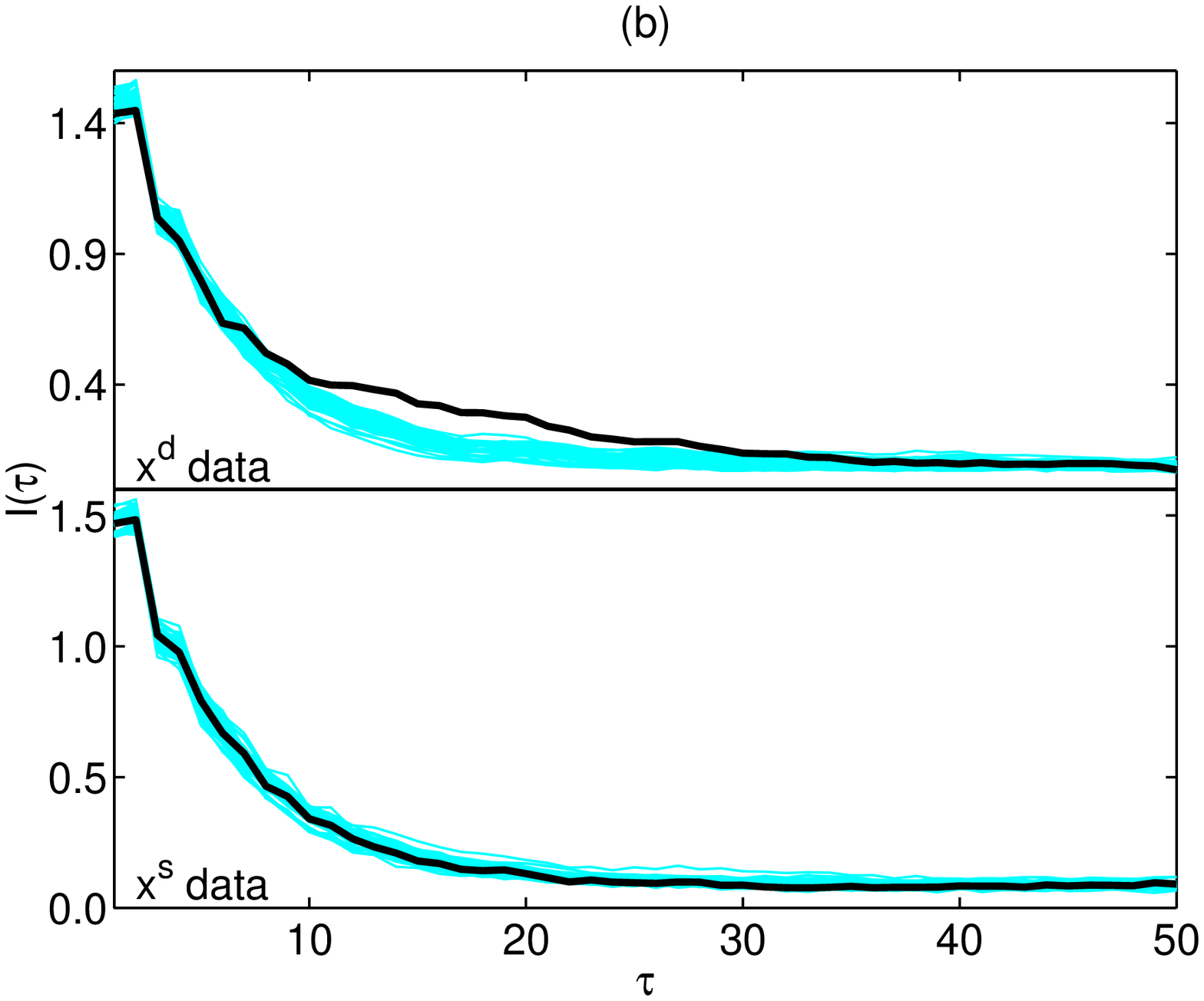,height=10cm,width=10cm}

\clearpage
\centerline{Figure 4c}
\psfig{file=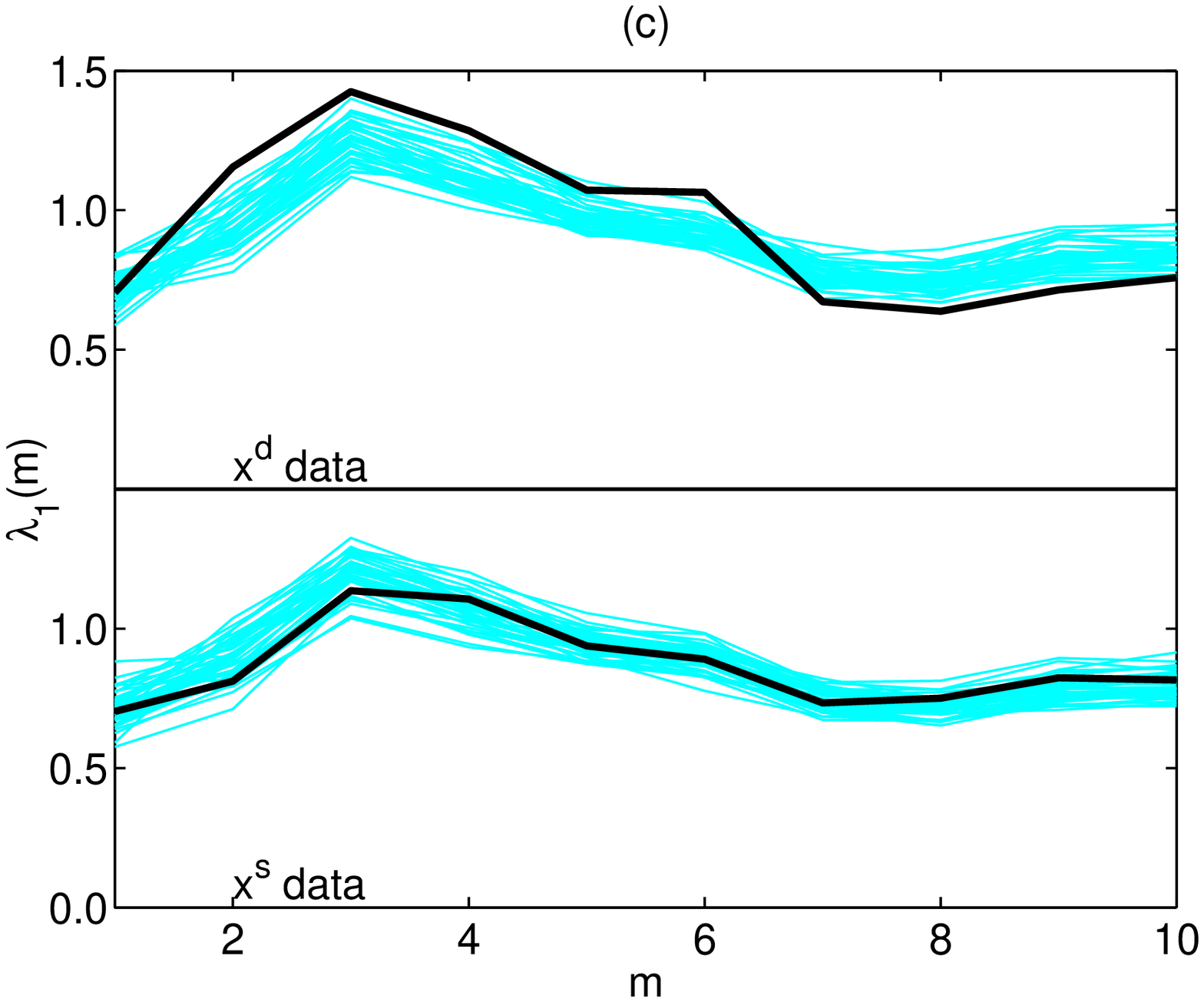,height=10cm,width=10cm}

\clearpage
\centerline{Figure 4d}
\psfig{file=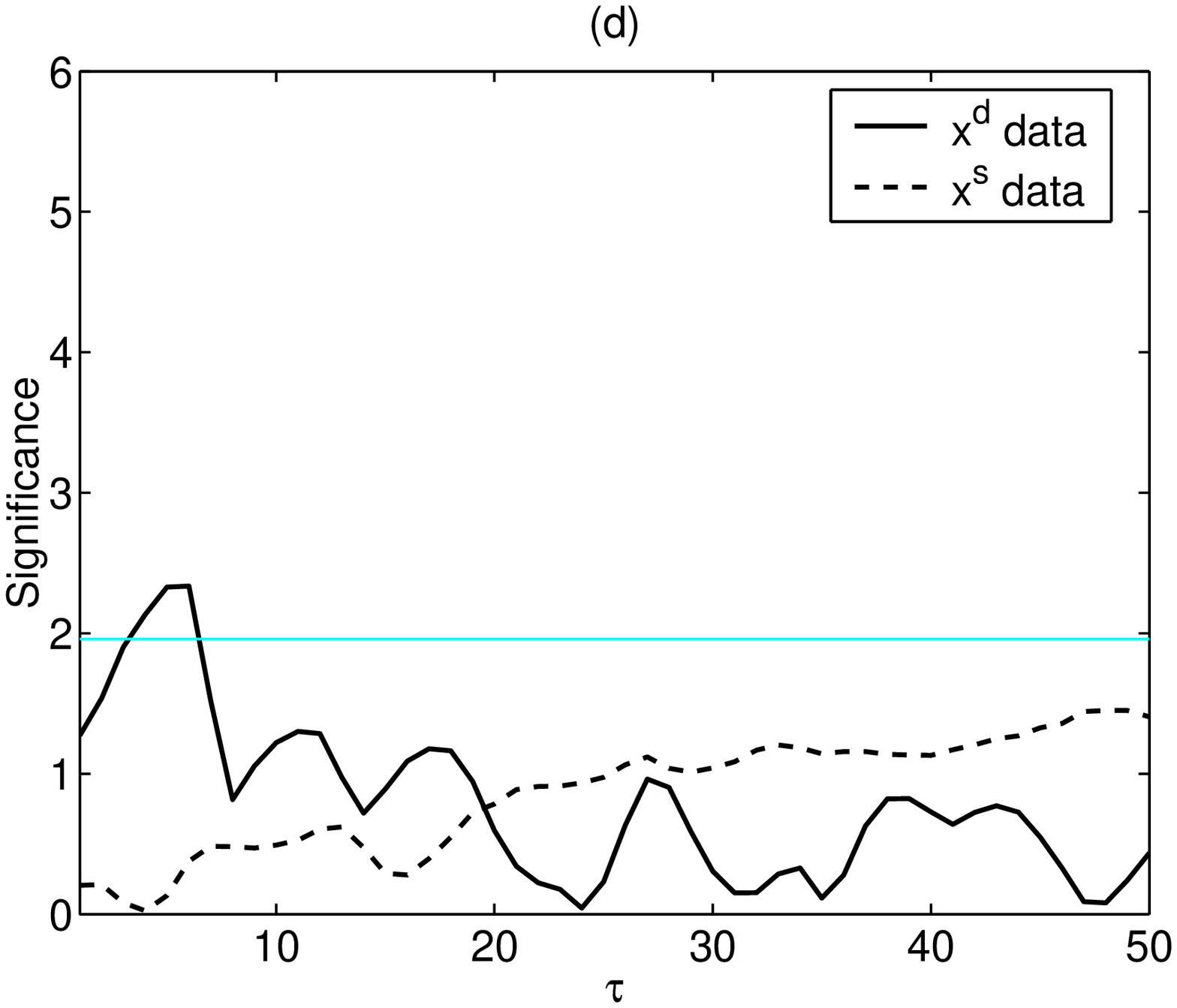,height=10cm,width=10cm}

\clearpage
\centerline{Figure 4e}
\psfig{file=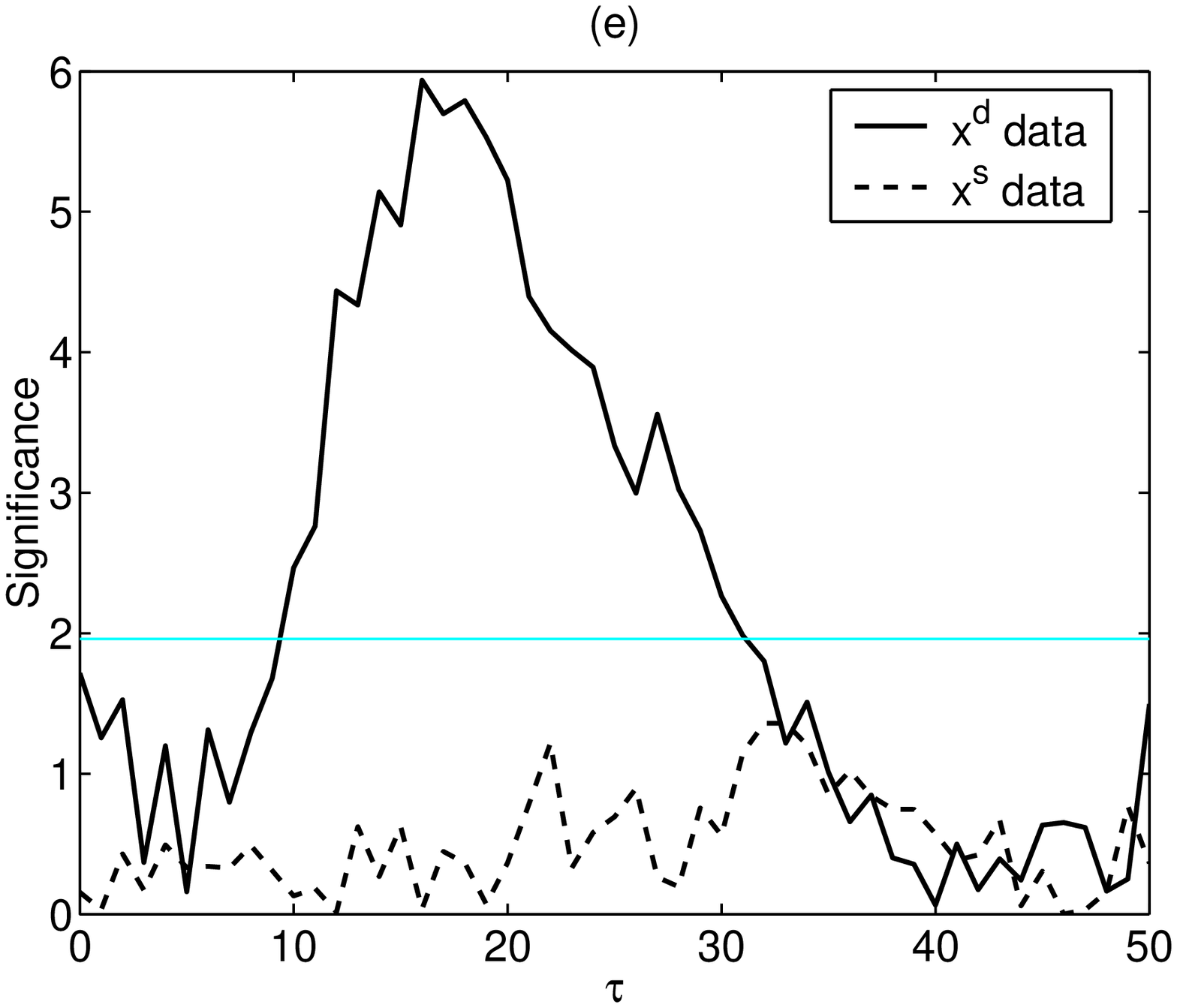,height=10cm,width=10cm}

\clearpage
\centerline{Figure 4f}
\psfig{file=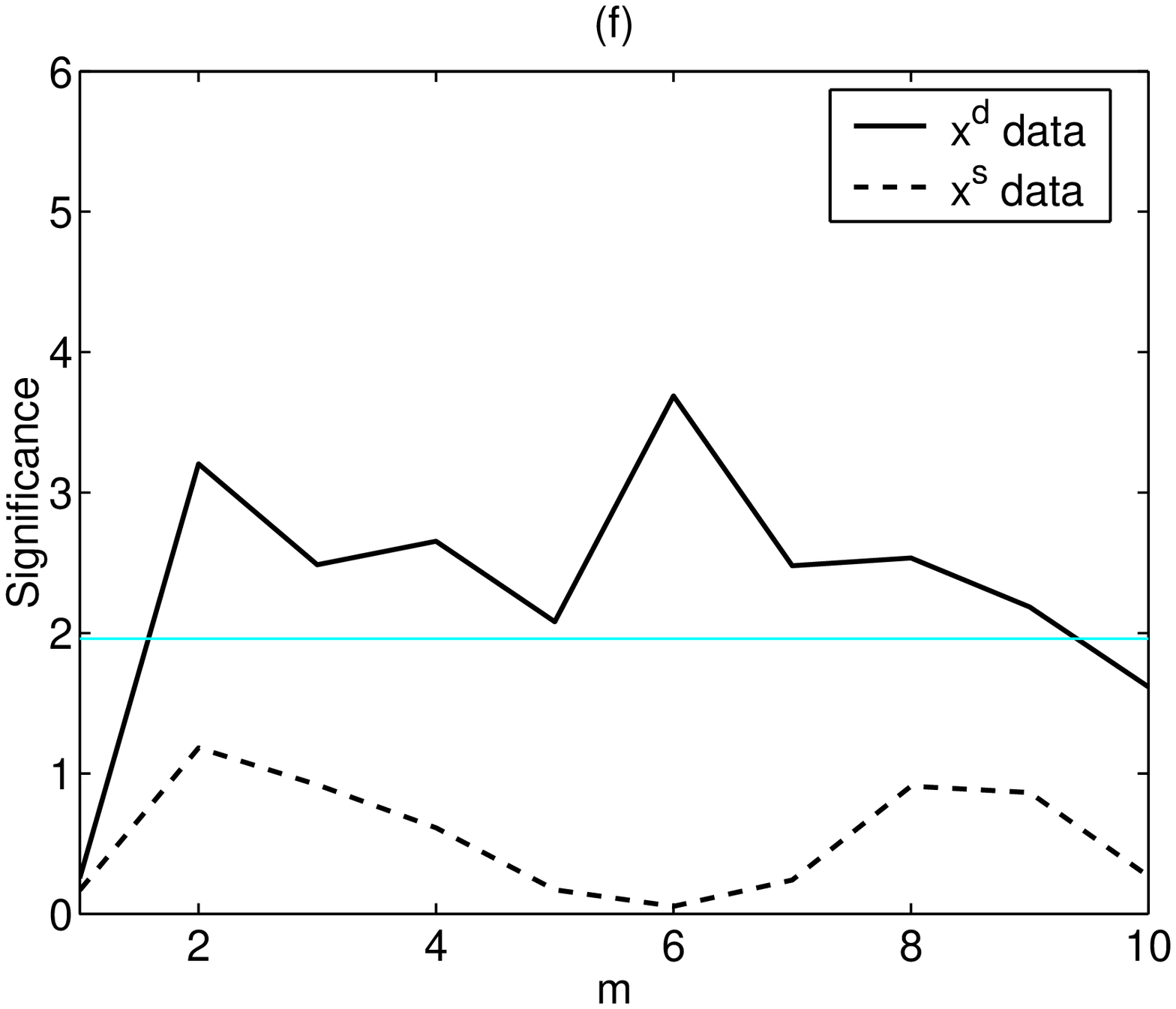,height=10cm,width=10cm}

\clearpage
\centerline{Figure 5a}
\psfig{file=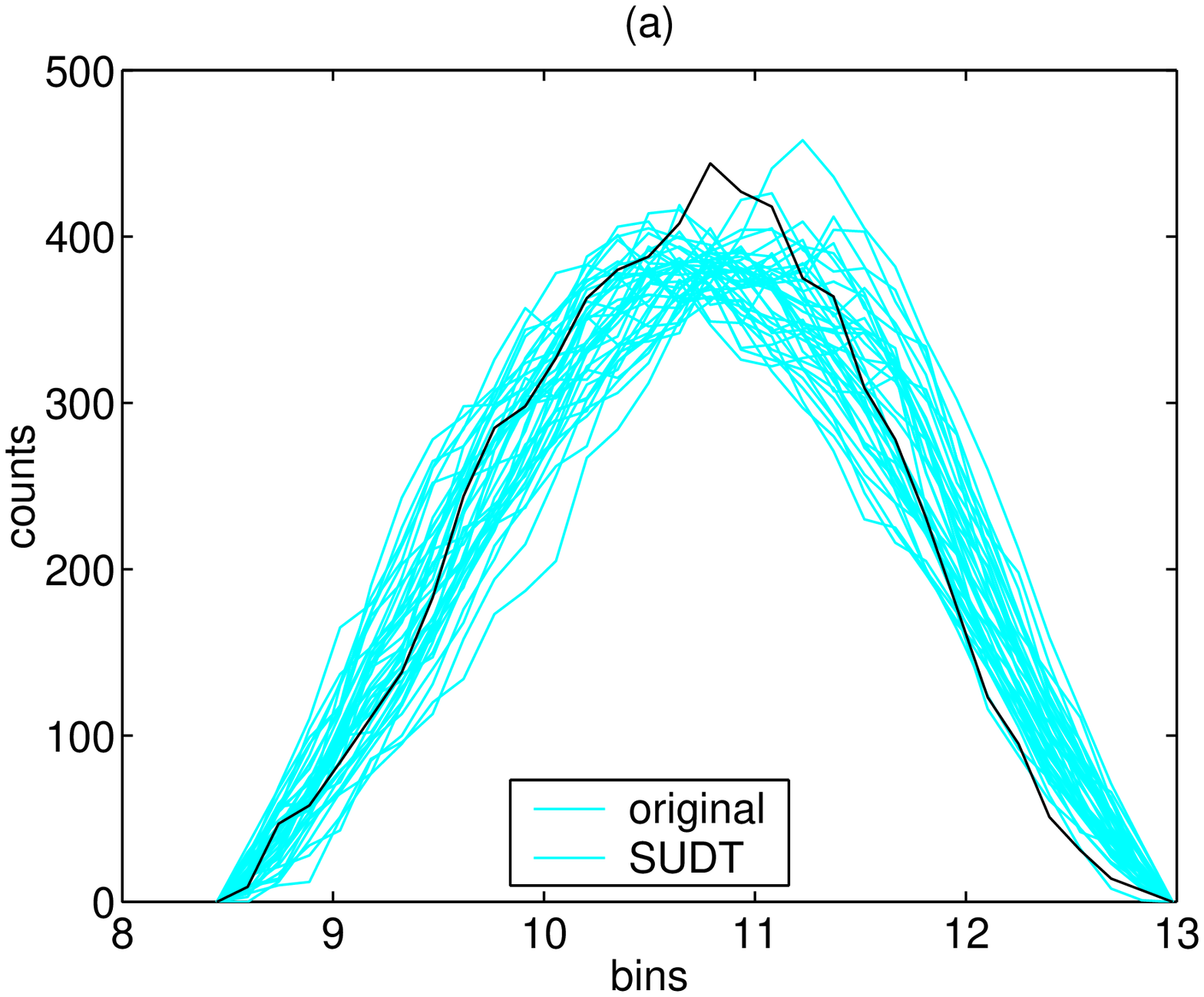,height=10cm,width=10cm}

\clearpage
\centerline{Figure 5b}
\psfig{file=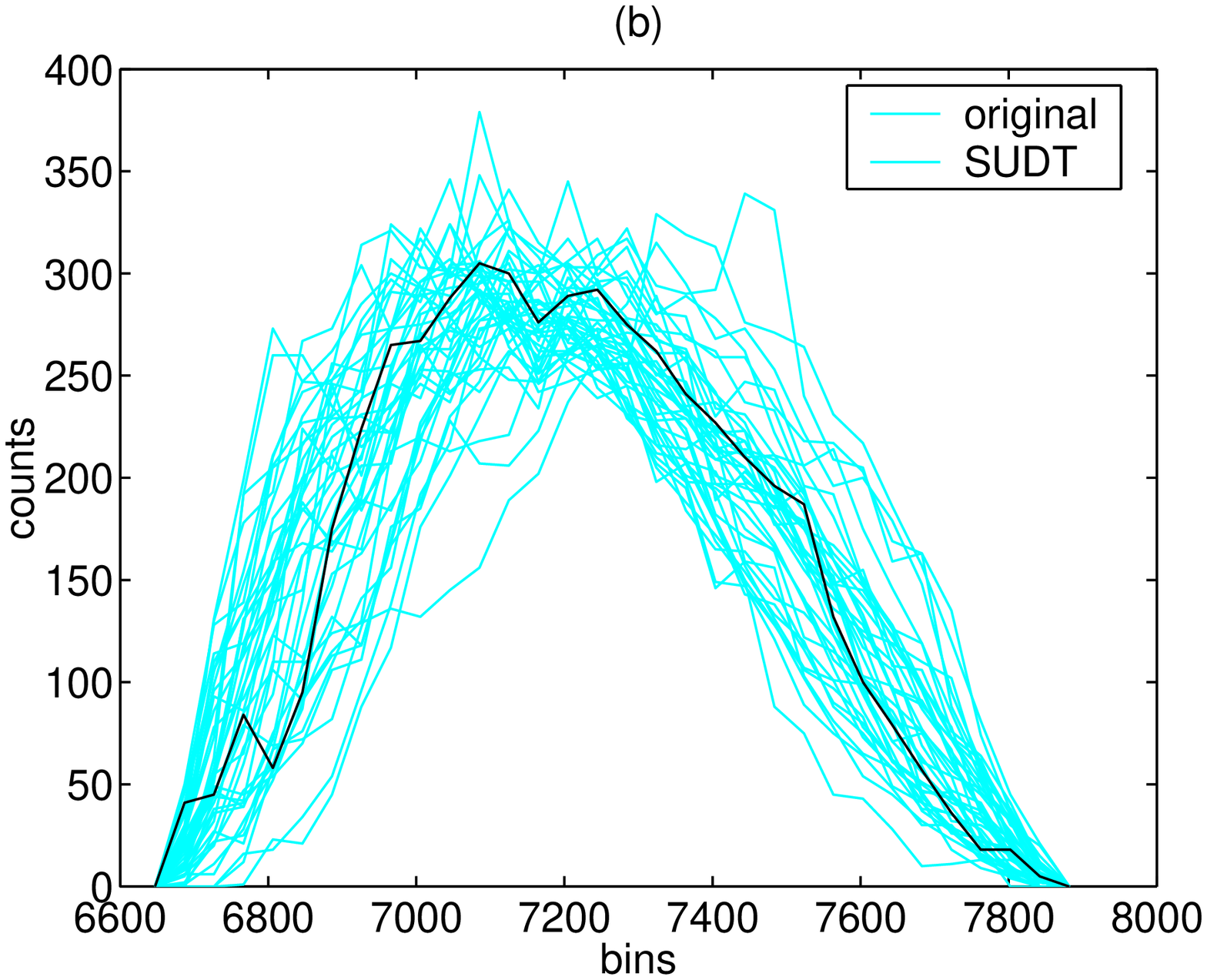,height=10cm,width=10cm}

\clearpage
\centerline{Figure 5c}
\psfig{file=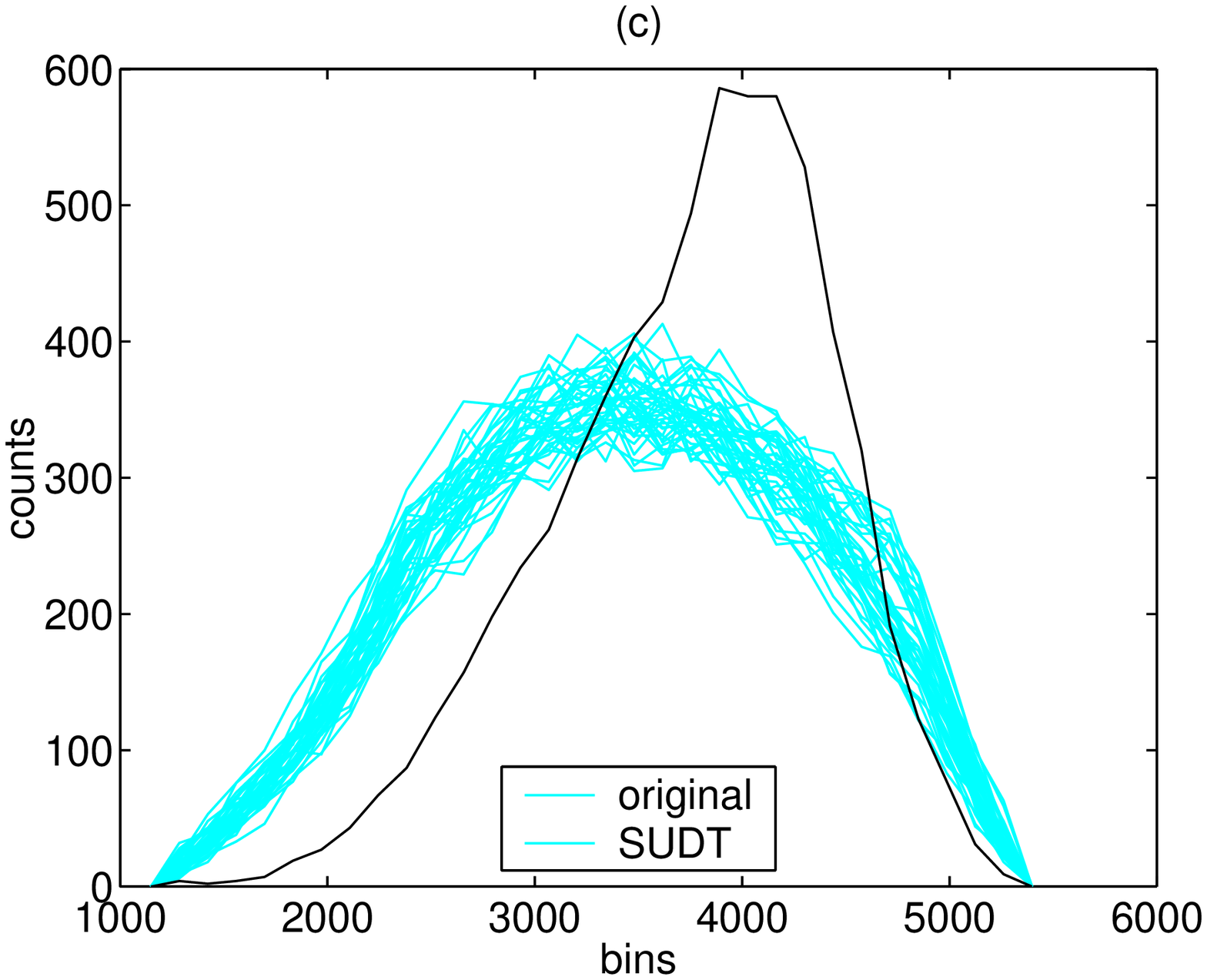,height=10cm,width=10cm}

\clearpage
\centerline{Figure 6a}
\psfig{file=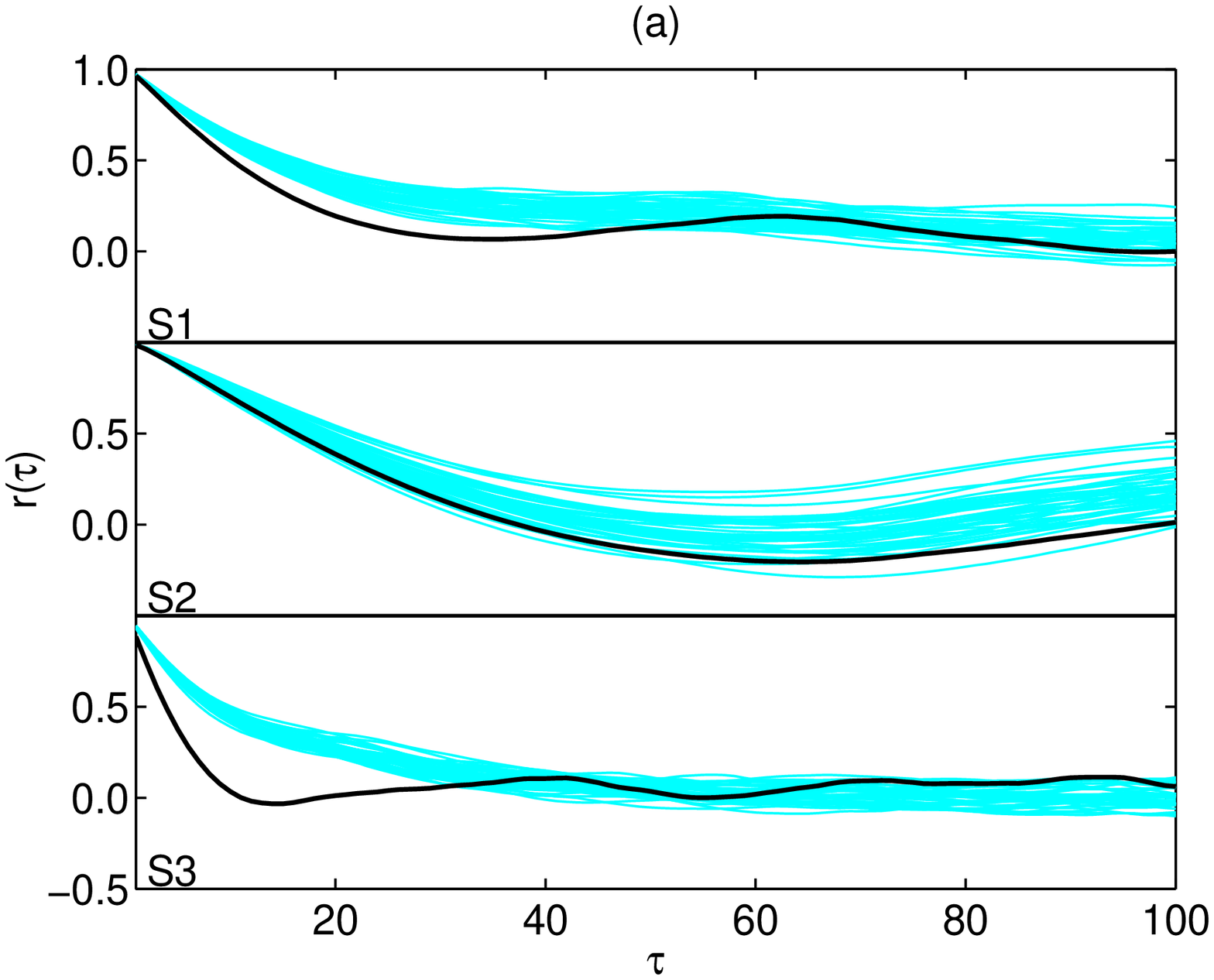,height=10cm,width=10cm}

\clearpage
\centerline{Figure 6b}
\psfig{file=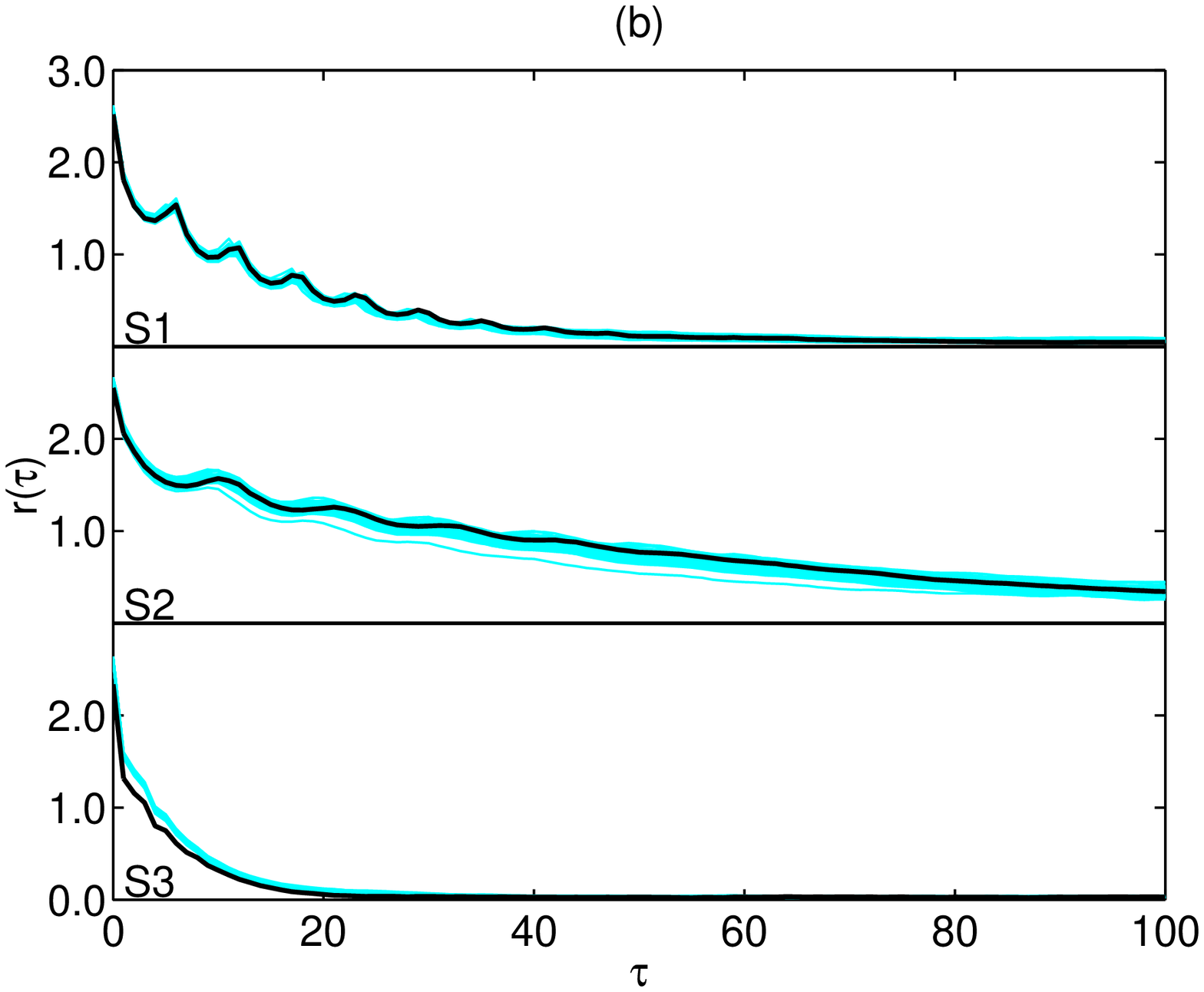,height=10cm,width=10cm}

\clearpage
\centerline{Figure 6c}
\psfig{file=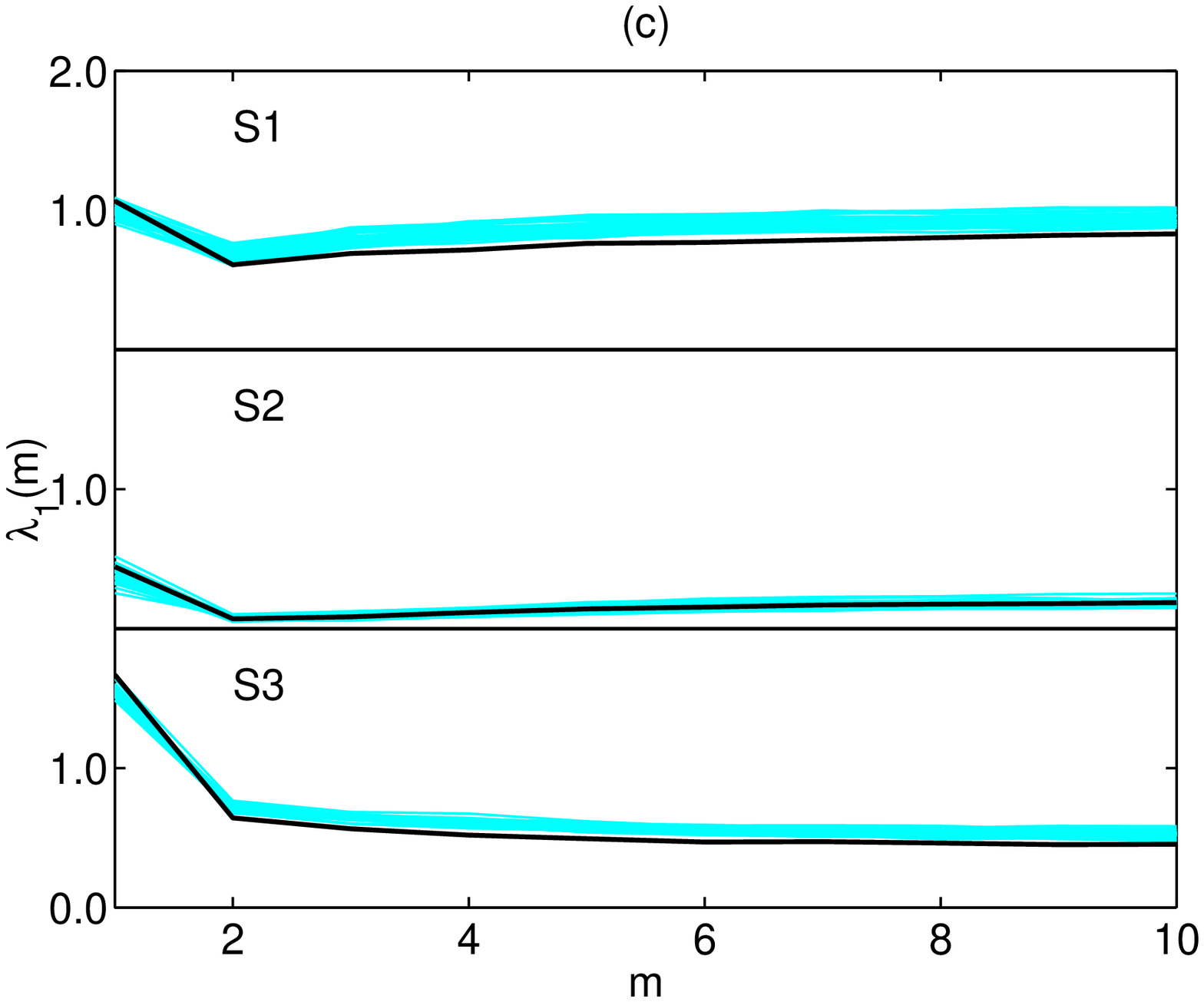,height=10cm,width=10cm}

\clearpage
\centerline{Figure 6d}
\psfig{file=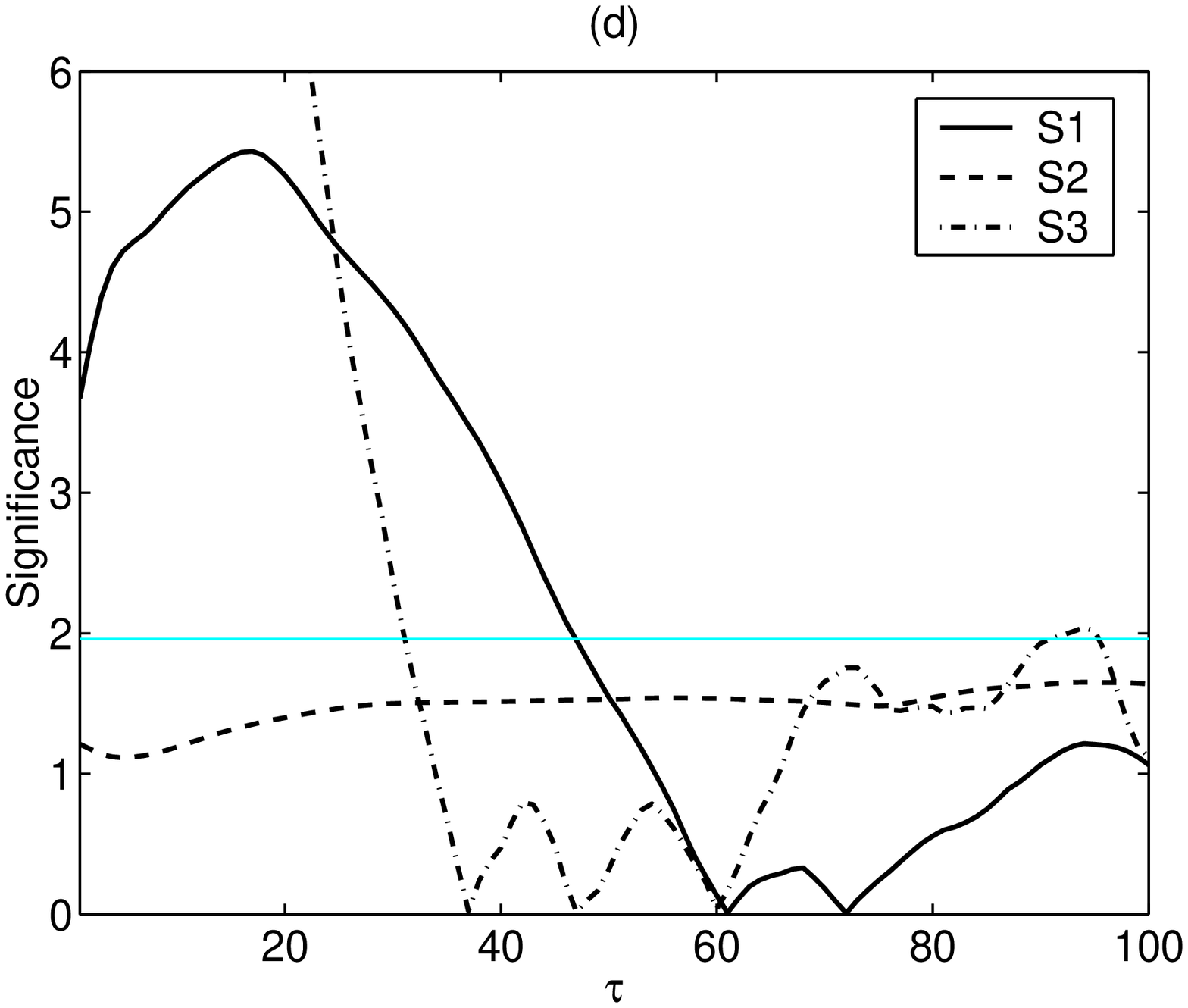,height=10cm,width=10cm}

\clearpage
\centerline{Figure 6e}
\psfig{file=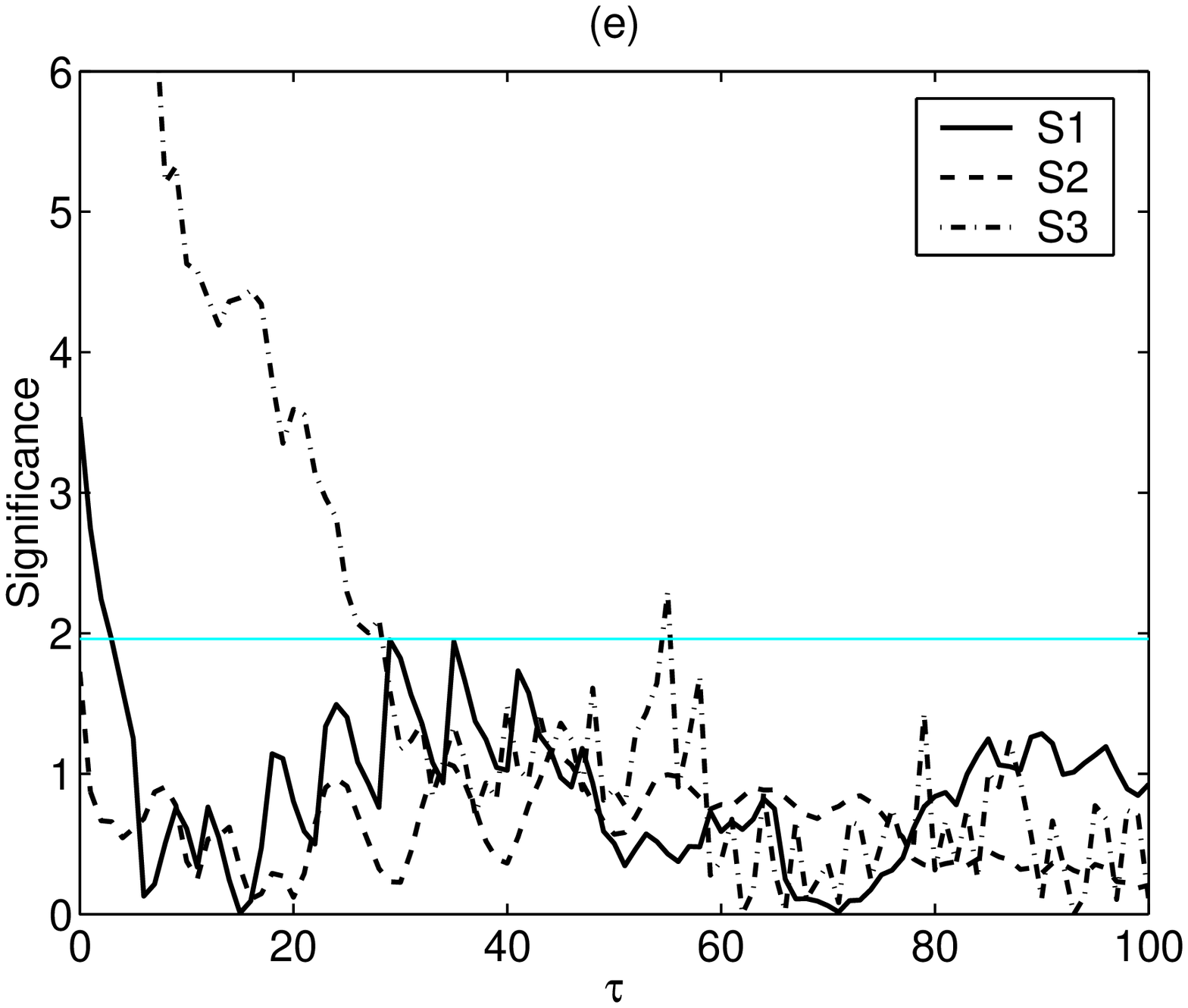,height=10cm,width=10cm}
	
\clearpage
\centerline{Figure 6f}
\psfig{file=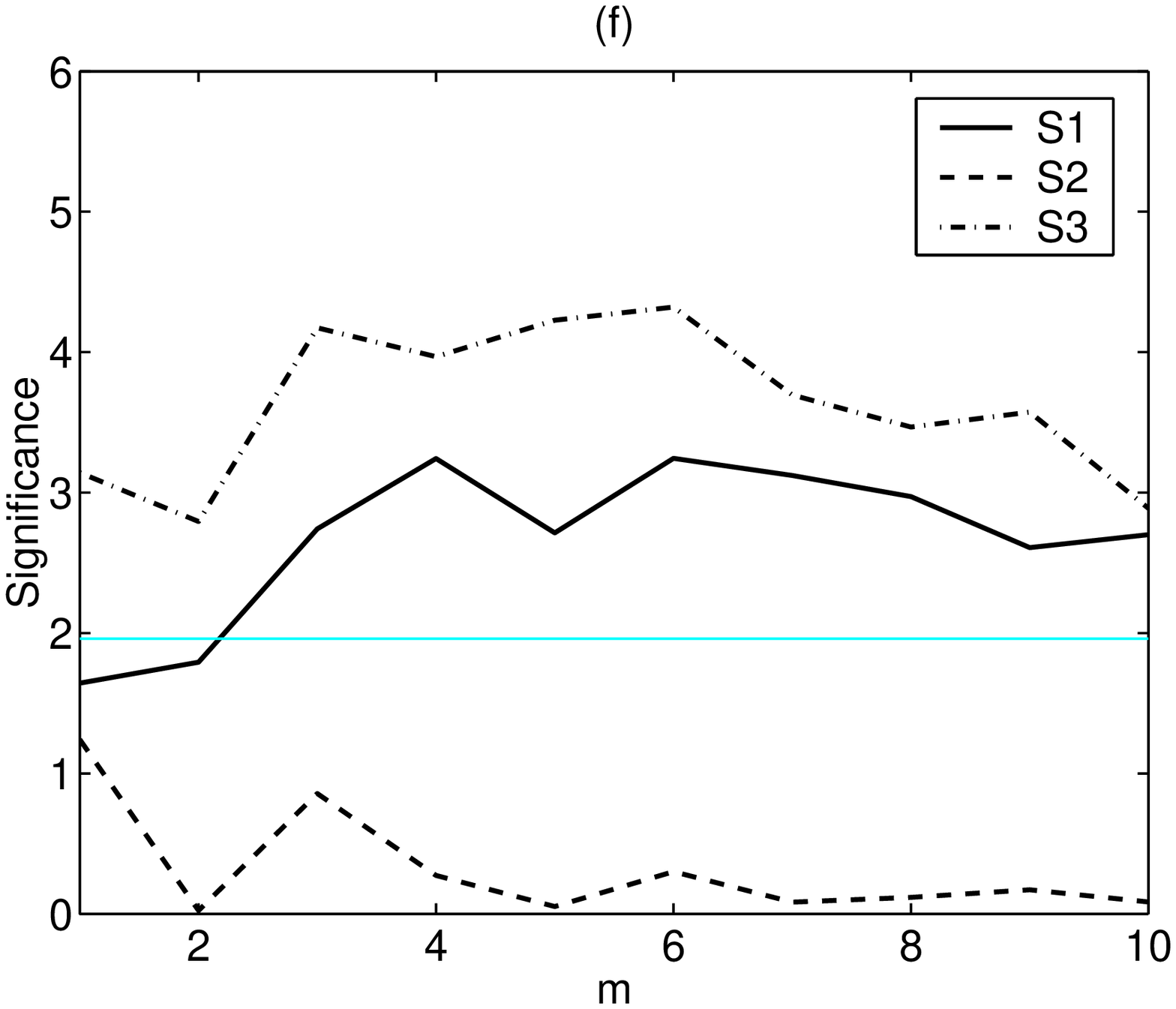,height=10cm,width=10cm}

\end{document}